\documentclass[a4paper,11pt]{article}
\usepackage{jcappub}
\usepackage{lineno}
\usepackage{graphicx}
\usepackage{subcaption}
\usepackage{array}      
\usepackage{multirow}   
\usepackage{booktabs}

\usepackage{array}      
\usepackage{multirow}   
\usepackage{booktabs}   

\usepackage{soul} 
\usepackage{xcolor} 

\newcommand{\beq}{\begin{equation}}
\newcommand{\eeq}{\end{equation}}
\newcommand{\bea}{\begin{eqnarray}}
\newcommand{\eea}{\end{eqnarray}}

\title{\boldmath The neutrino magnetic moment and supernovae with strong magnetic fields}

\author[a]{Cl\'ement Ehrhardt,}\note{Corresponding author.}
\author[b]{Maria Cristina Volpe,}\note{Corresponding author.}
\affiliation[a]{GSI Helmholtz-Zentrum für Schwerionenforschung GmbH, \\ Darmstadt, 64291, Germany}
\affiliation[b]{CNRS, Université Paris Cité, Astroparticule et Cosmologie,\\ F-75013 Paris, France}
\emailAdd{c.ehrhardt@gsi.de,volpe@apc.in2p3.fr,miguel.a.aloy@uv.es,martin.obergaulinger@uv.es}
\author[c,d]{Miguel \'Angel Aloy,}
\author[c,d]{Martin Obergaulinger}
\affiliation[c]{Departament d’Astronomia i Astrofísica, Universitat de Val\`encia, Av. Vincent Andrés Estellés, 19, 46100 Burjassot,
Spain}
\affiliation[d]{Observatori Astron\`omic, Universitat de Val\`encia, 46980 Paterna, Spain}

\abstract{
Despite several decades of searches, the neutrino magnetic moment remains unknown. 
If it is large, the neutrino evolution can be affected by its coupling to strong magnetic fields like the ones present in at least some core-collapse supernovae.
We explore these effects for Majorana neutrinos. We first discuss a schematic model often employed and show the impact of resonant spin-flavor precession on the neutrino signal for a future galactic supernova. Then, we consider detailed simulations of a neutrino-driven supernova with 17 ${\rm M}_{\odot}$ leaving a neutron star and a magnetorotational supernova with 26 ${\rm M}_{\odot}$ with a black-hole remnant.  We present predictions for the supernova neutrino fluxes and the event rates in presence of a non-zero neutrino magnetic moment, for the running JUNO and upcoming Hyper-Kamiokande experiments.
Our results show a significant impact for transition magnetic moments of $10^{-14} ~\mu_B$ during the neutronization burst and the accretion phase for the neutrino driven supernova, and as small as $10^{-15} ~\mu_B$ at the early times for the magnetorotational supernova.}
\begin{document}
\maketitle
\flushbottom

\section{Introduction} 
Neutrinos have revealed many of their mysteries. Neutrino mixings and mass-squared differences have been determined with precision by solar, reactor and accelerator experiments \cite{ParticleDataGroup:2024cfk}. The most precise determination of the solar mixing parameters was recently provided with the first analysis release by the Jiangmen Underground Neutrino Observatory (JUNO) experiment \cite{JUNO:2025gmd}. On the other hand, neutrinos keep being unique astrophysical and cosmological probes.  
The measurement of neutrinos from the carbon-nitrogen-oxygen cycle, first suggested by H. Bethe back in 1939, seems to favor the high metallicity solar models \cite{BOREXINO:2023ygs}. The IceCube experiment opened the observational window in high energy neutrino astrophysics more than a decade ago \cite{IceCube:2013cdw}, while the KM3NeT Collaboration detected the most energetic neutrino ever \cite{KM3NeT:2025npi}. 

Still, many neutrino properties remain unknown.  
The analysis of the KATRIN experiment furnished the upper limit of 0.45 eV at 90 $\%$ C.L. for the effective electron neutrino mass \cite{KATRIN:2024cdt}. Cosmology provides important \textit{albeit} model-dependent information on the sum of the neutrino masses \cite{ParticleDataGroup:2024cfk}. Ongoing and upcoming experiments will measure the neutrino mass ordering, and CP violation in the neutrino sector. For the former normal mass ordering (NO) seems to be favored over the inverted one (IO) \cite{Wang}; while for the latter CP violation is weakly preferred (NO) over CP conservation  \cite{Capozzi:2025wyn}.  Determining if neutrinos are Dirac or Majorana particles would represent a breakthrough. 
Challenging open questions also include understanding how neutrinos change flavor in dense environments including core-collapse supernovae and binary neutron star mergers remnants. Indeed the neutrino-neutrino interaction makes neutrino propagation in theses sites an entangled weakly interacting many-body problem (for reviews see e.g. \cite{Volpe:2015rla,Tamborra:2020cul,Volpe:2023met,Johns:2025mlm}). 

Among the unknowns are neutrino electromagnetic properties, including the neutrino magnetic dipole moment that neutrinos acquire due to quantum loop effects \cite{Giunti:2014ixa}. While in the Standard Model (minimally extended to include the neutrino mass) neutrinos have a small magnetic moment,  
beyond the Standard Model physics could produce large values of $\mu_{\nu}$ (see e.g. \cite{Babu:2020ivd, Jana:2022tsa}). 
Interestingly, electromagnetic properties depend on the neutrino nature. While Dirac neutrinos have both diagonal and off-diagonal ({\it albeit} suppressed) transition magnetic moments, Majorana neutrinos only have the latter. This produces a different impact in astrophysical sites when considering observable effects of resonant spin-flavor precession (RSFP) due to the coupling of the magnetic moment to magnetic fields. 

The detection of the neutrino magnetic moment is a longstanding quest that goes back to 1957 when Reines and Cowan's measurement using reactors set the first upper limit \cite{Cowan:1957pp}. 
Indeed, a non-zero $\mu$ introduces extra contributions to a variety of microscopic processes, such as neutrino-electron scattering, or plasmon decay important for the cooling of stars as, e.g., red giants. Numerous reactor- and accelerator-based experiments have been searching for large magnetic moments. Tight constraints are also obtained thanks to astrophysical sources, including the Sun, red giants or core-collapse supernovae. Table \ref{tab:mulimits} summarizes several upper bounds on $\mu$ determined using a variety of neutrino sources.
It is to be noted that theoretically strong model-independent bounds can be put for Dirac neutrinos \cite{Bell:2005kz}, whereas for Majorana neutrinos these are much weaker. Thus the detection of a large magnetic moment could indicate that neutrinos are Majorana particles \cite{Bell:2006wi}. Theoretical aspects of RSFP were also discussed in ref.~\cite{Dvornikov:2025mdu}, and including many-body correlations in ref.~\cite{Pehlivan:2014zua}.

Historically, the neutrino magnetic moment was considered as a possible solution of the solar neutrino problem, due to its coupling to solar magnetic fields in the convective zone that would produce RSFP (see e.g. \cite{Cisneros:1970nq,Okun:1986na,Balantekin:1990jg,Beacom:1999wx,Raghavan:1991em}). It is nowadays known that, even if we take observational upper limits for the solar magnetic field, magnetic moments beyond current bounds would be required to produce sizable spin-flavor effects \cite{Balantekin:2004tk}. 
The impact of the neutrino magnetic moment was investigated in cosmology \cite{Grohs:2023xwa,Vassh:2015yza,Carenza:2022ngg}, in relation with SN1987A, future core-collapse supernovae
(see e.g. \cite{Barbieri:1988nh,Barbieri:1988xw,Balantekin:2007xq,Ando:2002sk,Leurer:1988av,Abbar:2020ggq,Jana:2022tsa,deGouvea:2012hg,Abbar:2024nhz,Sasaki:2021bvu,Kharlanov:2020cti,Pehlivan:2014zua,Volpe:2007qx,Athar:1995cx}), $r$-process nucleosynthesis (see e.g. \cite{Balantekin:2007xq} and intermediate mass-stars \cite{Mori:2020niw}. RSFP was also investigated for the first time, on the diffuse supernova neutrino background in ref.~\cite{Volpe:2007qx},
and recently in ref.~\cite{Manno:2026jrg}.

\begin{table}\label{tab:limits}
    \centering
\begin{tabular}{lccc}
\toprule
Upper bounds [in $\mu_{B}$] & CL  & Experiment & Reference \\
\midrule
 $~~~~~~~10^{-9}$ & & Reines, Cowan    & \cite{Cowan:1957pp} \\
 $2.2  \times 10^{-10}$&  & TEXONO & \cite{TEXONO:2009knm} \\
 $ 2.9 \times 10^{-11} $ & 90 $\%$  & GEMMA & \cite{Beda:2013mta} \\
 $~~9 \times 10^{-11}$ & 90 $\%$  &
 MUNU & \cite{MUNU:2005xnz} \\ 
  $6.8 \times 10^{-10}  $ & 90 $\%$ & LSND &  \cite{LSND:2001akn} \\
 $7.5 \times 10^{-11}$ & 90 $\%$  & CONUS & \cite{CONUS:2022qbb} \\
 $ 6.4 \times 10^{-12}$ & 90 $\%$ & XENONnT & \cite{XENON:2022ltv} \\ 
  $ 2.8 \times 10^{-11}$ & 90 $\%$ & Borexino &\cite{Borexino:2017fbd} \\
  $2$-$8 \times 10^{-11}$ & &  SN1987A & \cite{Barbieri:1988nh} \\
$ ~~~3 \times 10^{-12}$ & & Globular cluster stars & ~\cite{Raffelt:1990pj} \\
  $1.2 \times 10^{-12}$ & 95 $\%$ & Red Giant Branch Tip & \cite{Capozzi:2020cbu}\\
\midrule
$~~~~10^{-14}$  & & Theory (Dirac $\nu$) & \cite{Bell:2005kz}\\
\bottomrule
\end{tabular}
\caption{Limits on the neutrino magnetic moment from experiments using neutrinos from accelerators, reactors, and from astrophysical sources, in particular the Sun, SN1987A and red giants. For Dirac neutrinos, a model-independent theoretical bound is also shown.} \label{tab:mulimits}
\end{table}

Interestingly, simulations of rapidly rotating stellar core collapse and proto-neutron (PNS) star evolution produce local or internal magnetic fields reaching $10^{15}$-$10^{16} $ Gauss. Flux compression and differential rotation winding amplify the field during and after collapse, while the magnetorotational instability (MRI) and MRI-driven dynamo action can generate strong turbulent and large-scale magnetic components \cite{Shibata:2006hr,Sawai:2013exa,Mosta:2015ucs,Obergaulinger:2021omt,Reboul-Salze:2020mnw}.
Magnetorotational supernovae only represent a few percent of supernovae, the majority of which is neutrino driven (see e.g. \cite{Janka:2017vlw}). Associated with the most energetic events, like superluminous supernovae \cite{AguileraDena:2018},
magnetorotational supernovae can also be a source of $r$-process elements \cite{Winteler:2012hu,Reichert:2024vyd} and of gravitational waves. Ref.~\cite{Manno:2026ikk} investigated the impact of neutrino-antineutrino conversions due to transition magnetic moments in a magnetorotational supernova with zero-age main-sequence mass of 13 $M_{\odot}$, considering the directional dependence of the neutrino emission.
  
In this work, we explore the impact of the neutrino magnetic moment for Majorana neutrinos including 3$\nu$ flavors. First we consider a schematic model, often employed in the literature (see e.g. \cite{Akhmedov:2003fu,Jana:2022tsa}), where the radial matter density and the magnetic field profiles are given by power-laws. We use the model to present the RSFP mechanisms and numerically determine matter eigenvalues, neutrino probabilities and fluxes.  
We then turn to the main focus on the paper which is to investigate the impact of the neutrino magnetic dipole moment, based on detailed simulation results from two supernova models with rotation and strong magnetic fields. One star, with a zero-age main sequence mass of 17 ${\rm M}_{\odot}$ explodes driven by the standard neutrino-driven mechanism and leaves behind a neutron star, while the other one, for a zero-age main sequence mass of 26 ${\rm M}_{\odot}$, produces a magnetorotational explosion and a collapse of the PNS to a black-hole.
These models are full three-dimensional versions of previous  two-dimensional simulations \cite{Obergaulinger:2021omt}. A detailed description of the dynamics of the new models will be provided in Ref.~\cite{Aloy-Obergaulinger}. 
We investigate the neutrino evolution and spectral modifications induced by sizable neutrino transition magnetic moments. By considering the time-dependent neutrino emission and stellar profiles, we present 
supernova neutrino fluxes and the expected event rate in the running JUNO and upcoming Hyper-Kamiokande (HK) experiments (starting in 2028), in absence and in presence of large magnetic moments. We show that significant effects, especially at early times in both models, can be induced even for magnetic moment values well below current experimental bounds. 

The manuscript is structured as follows.
Section \ref{Sect:theory} sets the 3$\nu$ theoretical framework for RSFP and the conditions for the spin-flavor resonances. Section \ref{Sect:schematic} presents the schematic model, and the corresponding numerical results on the matter eigenvalues, neutrino probabilities and fluxes.
Section \ref{Sect:simulations} describes the three-dimensional neutrino-driven and magnetorotational supernova simulations considered in this study.
The numerical results using the detailed supernova models and including neutrino-antineutrino conversions are presented in Section \ref{Sect:results}, both for the neutrino fluxes and rates in JUNO and HK. Section \ref{Sect:conclusions} gathers our Conclusions.

\section{Theoretical framework}
\label{Sect:theory}

In the Standard Model, 
neutrinos can acquire a magnetic moment from quantum loop effects 
 (see e.g., \cite{Giunti:2014ixa} for a comprehensive review). For Dirac neutrinos the diagonal magnetic moment in the mass basis reads 
\begin{equation}
 \mu^D_{ii} \simeq
{{3 e G_F m_{i}} \over{8 \sqrt{2} \pi^2 }} \Big[
1 + O\Big( {{m_l} \over {m_{\rm W}}} \Big)^2
\Big]
\simeq 3.2 \times 10^{-19} \Big({m_{\nu}\over
1 ~{\rm eV}} \Big) \mu_B \ , 
\end{equation}
where $\mu_B$ is the Bohr magneton, $m_l$ and $m_{\rm W}$ are the charged lepton and the W-boson mass respectively. 
Their ratio squared is at most $5 \times 10^{-4}$.
The transition magnetic moments are given by
\begin{equation}
 \mu^D_{ij}  \simeq
{{-3 e G_F} \over{32 \sqrt{2} \pi^2 }} (m_i + m_j) \sum_{l = e, \mu, \tau} U^*_{li} U_{l j} {m_l^2 \over m_{\rm W}^2} \ , 
\end{equation}
where $U$ is the Pontecorvo-Maki-Nakagawa-Sakata (PMNS) unitary matrix.
The transition magnetic moments between different neutrino mass eigenstates are ``GIM suppressed" in analogy with the suppression of the flavor-changing one-loop contributions for hadronic processes in the Glashow-Iliopoulos–Maiani (GIM) mechanism \cite{Glashow:1970gm}. For Majorana neutrinos, the magnetic moment matrix is totally antisymmetric, i.e.
 $ \mu_{ij}^M = - \mu_{ji}^M $ for $i,j = 1, 2, 3$ so that only transition magnetic moments are allowed.
These read \cite{Giunti:2014ixa}
\begin{equation}
 \mu^M_{ij}  \simeq
{{-3 i e G_F} \over{16  \sqrt{2} \pi^2}} (m_i + m_j) \sum_{l = e, \mu, \tau} {\rm Im}[U^*_{li} U_{l j}] {m_l^2 \over m_{\rm W}^2} \ , 
\end{equation}
where here the mixing matrix can include the Majorana phases. As in the Dirac case the transition magnetic moments are suppressed by the ``GIM mechanism". 

 An interesting consequence of the neutrino magnetic moment coupling to stellar magnetic fields is the spin-flavor mechanism where both neutrino helicity and flavor can change. If neutrinos are Dirac particles, the helicity flip changes a neutrino into a sterile one. On the contrary, if neutrinos are Majorana, helicity flipping only involves active neutrinos.
 
\subsection{Neutrino evolution equations in presence of transition magnetic moments}
 When neutrinos are produced in astrophysical or cosmological environments, they interact with the particles composing the medium, e.g. electrons, protons, neutrons, and the neutrinos or antineutrinos, through standard model charged-current (CC) and neutral-current (NC) interactions. In the flavor basis, the neutrino evolution equations for the neutrino flavor states $\vert \nu_{\alpha} (t) \rangle$ read\footnote{From now on we will use $\hbar=c=1$.}
\beq\label{e:Schro}
i {d  \over dt} \vert \nu_{\alpha} (t) \rangle = {\cal H}  \vert \nu_{\alpha} (t) \rangle \, ,
\eeq
with $\vert  \nu_{\alpha} (0) \rangle = \vert  \nu_{\alpha} \rangle$ as initial condition. 
The neutrino flavor basis is related to the mass basis through the relation 
\beq \label{eq:VO}
\vert \nu_{\alpha} \rangle = \sum_{i} U^{*}_{\alpha i } ~ \vert \nu_{i} \rangle \, ,   
\eeq
with  the flavor and mass indices $\alpha = e, \mu, \tau, \dots $ and $i = 1, 2, 3, ..., N$  respectively, $N$ being the number of neutrino families.
 
The neutrino Hamiltonian receives different contributions 
\beq\label{e:H}
{\cal H} = {\cal H}_{\rm vac}^f + {\cal H}_{\rm mat} + {\cal H}_{\rm \mu B} \, ,
\eeq
where the first is the vacuum term, the second comes from neutrino-matter interactions and the last term accounts for the coupling of the neutrino magnetic dipole moment to magnetic fields. 
The last contribution to the Hamiltonian introduces neutrino helicity flipping. As a consequence the $2N\times 2N$ neutrino Hamiltonian has both flavor and helicity structure (see e.g. \cite{Pehlivan:2014zua}). This is analogous to the case 
of neutrino helicity change from wrong-helicity contributions due to the absolute neutrino mass, as first pointed out in ref.~\cite{Volpe:2013uxl}. 
Termed ``spin-coherence" \cite{Vlasenko:2013fja} or ``helicity-coherence" \cite{Serreau:2014cfa}, these mean-field contributions can also produce neutrino-antineutrino mixing in anisotropic media. 

It is necessary to emphasize that neutrino propagation in dense environments -- the early Universe, core-collapse supernovae, binary neutron star merger remnants and accretion-disks around black-holes -- is a weakly interacting many-body problem. In such media, the neutrino-neutrino interaction is sizable and makes neutrino evolution non-linear producing a variety of new phenomena (see \cite{Volpe:2023met,Tamborra:2020cul,Johns:2025mlm} for reviews). 

Currently,
the full neutrino quantum kinetic equations have been solved for the early Universe, since homogeneity and isotropy make the challenging numerical problem tractable \cite{Froustey:2020mcq,Bennett:2020zkv}. In contrast, their solution in astrophysical sites requires further work.   
It is to be noted that a few studies on the neutrino magnetic moment and including the $\nu\nu$ interaction were performed. Ref.~\cite{Abbar:2020ggq} considered the 'line model' for Majorana neutrinos, with fixed values for the astrophysical inputs, and discussed flavor equilibrium when neutrino-antineutrino conversions occur due to a non-zero neutrino magnetic moment. Based on the 'line' model, refs.~\cite{Sasaki:2023sza} and \cite{Sasaki:2021bvu} investigated spin-flavor precession for Dirac and Majorana neutrinos, with fixed inputs from neutrino-driven supernova simulations, and searched for possible matter suppression.
In ref.~\cite{Kharlanov:2020cti} the impact of a non-zero magnetic-moment was studied including standard and nonstandard $\nu\nu$ interactions.

However, determining the neutrino evolution in dense environments is still an open problem involving both conceptual and numerical challenges.  
At present, investigations range from self-consistent approaches \cite{Volpe:2013uxl,Fiorillo:2025npi} to numerical treatments that evolve the short scale modes called "fast" in boxes of tens of kms and search for asymptotic states like flavor equilibration \cite{Cornelius:2024zsb}. It is to be noted that a study with a two-beam model showed that fast modes do not necessarily produce flavor equilibration \cite{Abbar:2018beu}. 
Since the problem of the treatment and the impact of the neutrino-neutrino interaction is not yet solved, we postpone the inclusion of the neutrino-neutrino interaction to future investigations. 

In our investigation we adopt a full 3$\nu$ framework and the mean-field approximation  \cite{Volpe:2023met,Volpe:2015rla}.
In this case the evolution equations are\footnote{From now on we employ $\hbar = c = 1$.}
\begin{equation}\label{eq:eveq3nu}
i {d \Psi \over{dr}} = 
\left(
\begin{array}{cc}
{\rm H}  & {\rm B M}  \\
{\rm B M}^{\dagger} & {\rm \bar{H}}  \\
\end{array}
\right)  \Psi ~~~~~~
\Psi = 
\left(
\begin{array}{c}
\nu  \\
\bar{\nu}  \\
\end{array}
\right) \ , 
\end{equation}
with the neutrino Hamiltonian
\begin{equation}\label{eq:H}
{\rm H} = {\rm H}_{\rm vac}  + {\rm V}(r) \ , 
\end{equation}
including the vacuum and the matter terms. 
In \eqref{eq:eveq3nu} ${\rm B}$ is the magnetic field strength,
${\rm M}$ the magnetic moment matrix and
$\Psi$ a vector of neutrino and antineutrino amplitudes. 
The ${\rm \bar{H}}$ Hamiltonian is for antineutrinos. 
The off-diagonal terms in \eqref{eq:eveq3nu} correspond to the neutrino magnetic moment coupling to the magnetic fields that introduces mixing between the neutrino and the antineutrino sectors. Such mixing can induce RSFP mechanisms, that we will also refer to as
resonant neutrino-antineutrino flavor conversions.

Explicitly the vacuum Hamiltonian in \eqref{eq:eveq3nu} is 
\begin{equation}\label{eq:H3vac}
{\rm H}_{\rm vac} = {1 \over {2E}} \left[ U \left(
\begin{array}{ccc}
0 & 0 & 0 \\
0 & \Delta m^2_{21} & 0 \\
0 & 0 & \Delta m^2_{31} \\
\end{array}
\right)
U^{\dagger} \right] \ , 
\end{equation}
with the mass-squared differences $\Delta m^2_{31} = m^2_{3} - m^2_{1}$ and $\Delta m^2_{21} = m^2_{2} - m^2_{1}$. In general the contribution due to the neutrino magnetic moment coupling to the magnetic field reads
\begin{equation}\label{eq:B3}
{\rm M} {\rm B}(r) = \left(
\begin{array}{ccc}
\mu_{ee} & \mu_{e\mu} &  \mu_{e\tau} \\
- \mu_{e\mu} & \mu_{\mu\mu}  &  \mu_{\mu\tau} \\
- \mu_{e\tau} &  - \mu_{\mu\tau} & \mu_{\tau\tau}  \\
\end{array}
\right)
{\rm B}_{\perp}(r) \ , 
\end{equation}
where ${\rm B}_{\perp}(r)$ is the transverse magnetic field. Since in the present work we focus on Majorana neutrinos, one has $\mu_{\alpha \alpha} = 0 $ for $\alpha = e, \mu, \tau$.
Moreover, for simplicity, we will assume that the transition magnetic moments are all equal, and denote them with $\mu$.

As for the neutrino-matter contribution, the mean-field approximation, where only coherent forward scattering is considered, gives \cite{Volpe:2023met}
\begin{equation}\label{eq:V}
{\rm V}(r) = {\rm diag}({\rm V}_e, {\rm V}_{\mu}, {\rm V}_{\tau}) \ , 
\end{equation}
with the  antineutrino potentials having a minus sign with respect to the neutrino ones.
The neutrino-matter potentials coming from CC and NC interactions are given by 
\begin{equation}\label{eq:pot}
   {\rm V}_e = {\rm V}_{\rm CC} + {\rm V}_{\rm NC} ~~~~{\rm V}_x = {\rm V}_{\rm NC} ~~ {\rm with } ~~{\rm V}_{\rm CC} = \sqrt{2} G_F n_e ~~ {\rm and }~~{\rm V}_{\rm NC} = -\sqrt{2} G_F n_n /2 \ , 
\end{equation}
with \(x=\mu,\tau\),
$G_F$ the Fermi coupling constant, $n_e$ and $n_n$ the electron and the neutron number densities. Note that, in a neutral medium, the potentials from neutral-current scattering of neutrino on electrons and on protons cancel each other.

\section{Resonant spin-flavor precession: a schematic model}
\label{Sect:schematic}

We discuss now the resonant spin-flavor conversion phenomena that can arise in astrophysical environments when strong magnetic fields are present. We use a schematic model often employed to study the impact of the magnetic moment on core-collapse supernova neutrinos (see e.g.
\cite{Akhmedov:1987nc, Jana:2022tsa,deGouvea:2013zp, Abbar:2020ggq}).
To discuss the resonance conditions we employ the  $2 \nu$ framework, for which
the evolution equations read\footnote{Here the explicit dependence on $r$ is omitted not to overburden the text.}
\begin{equation}\label{eq:eveq2nu}
i {d \over{dr}}
\left(
\begin{array}{c}
 \nu_{e } \\
\nu_{x }  \\
\bar{\nu}_{e}  \\
\bar{\nu}_{x}  \\
\end{array}
\right)
= \left(  \begin{array}{cccc}
 - {\Delta m^2 \over{4E}} 
\cos 2 \theta + {\rm V}_{e}   &   {\Delta m^2 \over{4E}} \sin 2 \theta &  0 & \mu {\rm B}_{\perp} \\
 {\Delta m^2 \over{4E}} \sin 2 \theta & {\Delta m^2 \over{4E}} \cos 2 \theta + {\rm V}_{x} &
- \mu {\rm B}_{\perp} & 0 \\  
0 &  - \mu {\rm B}_{\perp} & - {\Delta m^2 \over{4E}} \cos 2 \theta- {\rm V}_{e}  & {\Delta m^2 \over{4E}} \sin 2 \theta \\
    \mu {\rm B}_{\perp} & 0  & {\Delta m^2 \over{4E}} \sin 2 \theta  & {\Delta m^2 \over{4E}} \cos 2 \theta - {\rm V}_{x} \\
\end{array}\right)
 \left(
\begin{array}{c}
 \nu_{e}  \\
\nu_{x}  \\
\bar{\nu}_{e}   \\
\bar{\nu}_{x} \\
\end{array} 
\right) \ ,    
\end{equation}
where 
$\Delta m^2$ and $\theta$ are the neutrino squared mass-difference and mixing angle.

We present numerical results obtained for first the neutrino matter eigenvalues and probabilities, then the neutrino fluxes. 
The matter density profile is taken as a power law \cite{Akhmedov:1987nc}
\begin{equation}\label{eq:matprof}
    \rho(r) = \rho_0 \Big({r_0 \over r} \Big)^3   
     ~~ {\rm with} ~~
   \rho_0 = 10^{14}~ \Big({{\rm g} \over {\rm cm}^3} \Big)~~ {\rm and} ~~ r_0 = 10 ~ {\rm km} \ . 
\end{equation}
As in previous works based on the schematic model, the magnetic field strength profile we adopt is\footnote{In the literature, the value $k = 3$ was also employed. Note that the ${\rm B}_0$ value is larger than in previous works. This is done to account for the transverse magnetic field strength in the magnetorotational model A26, before black hole formation (see section \ref{sec:A26}).}
\begin{equation}\label{eq:Bprof}
   {\rm B}_{\perp}(r) = {\rm B}_0 \Big({r_0 \over r} \Big)^k   
     ~~ {\rm with} ~~ k = 2  ~~ {\rm and} ~~
  {\rm B}_0 = 10^{16} ~{\rm Gauss} ~~ r_0 = 10 ~ {\rm km} \ .
\end{equation}

We now present the numerical results for $3\nu$ flavors, on the neutrino matter eigenvalues and probabilities, obtained by using \eqref{eq:matprof} and \eqref{eq:Bprof}. We take for the three neutrino mixing angles $\theta_{12} = 33.6^{\circ}$, $\theta_{13} = 8.5^{\circ}$, $\theta_{23} = 48^{\circ} $,
and $\Delta m^2_{21} = 7.53 \times 10^{-5}~{\rm eV}^2$ and $|\Delta m^2_{32}| = 2.43 \times 10^{-3}~{\rm eV}^2 $ for the solar and the atmospheric neutrino mass-squared differences respectively \cite{ParticleDataGroup:2024cfk}. 
The Dirac CP violating phase is fixed at $\delta = 0^{\circ}$.

\subsection{Neutrino matter eigenvalues and probabilities}
Let us discuss the spin-flavor mechanisms.
From the neutrino Hamiltonian \eqref{eq:eveq2nu}, one can see that 
different resonance conditions are possible involving both the neutrino and the antineutrino sectors. These add to the ones associated with the established Mikheev-Smirnov-Wolfenstein (MSW) mechanism 
triggered by neutrino-electron forward scattering in matter \cite{Mikheev:1986wj,Wolfenstein:1977ue}. 

By using the relation 
\begin{equation}\label{es:pot2}
V_{\rm CC} + 2 V_{\rm NC} = {{\sqrt{2} G_F \rho}\over{m_N}}~ (2 Y_e - 1) \ , 
\end{equation}
one gets the following resonance conditions.
First the MSW high resonance\footnote{The $V_{\mu \tau}$ resonance, arising because of loop corrections \cite{Botella:1986wy}, does not play any role here and will be neglected. Note that $\nu_{\mu'} $ and $\nu_{\tau}'$ indicate the eigenstates when the $23$ sector is diagonal \cite{Dighe:1999bi}.} (MSW-H) triggers either 
$\nu_e \rightarrow \nu_{\tau'}$ (NO, $\Delta m^2_{31} > 0 $) or $\bar{\nu}_e \rightarrow \bar{\nu}_{\tau'}$ (IO, $\Delta m^2_{31} < 0 $) conversions. They take place if the following condition is fulfilled
\begin{equation}\label{eq:MSWH}
{\rm MSW-H: }~~~~~ \rho_{res} Y_e \simeq {{m_N \over{\sqrt{2} G_F}}{ {\Delta m^2_{31}} \over{2E}} \cos2\theta_{13}}  \ ,  
\end{equation}
where $\rho_{res}$ indicates the matter density at the resonance location, $m_N$ is the nucleon mass. 
The MSW low resonance (MSW-L) produces
$\nu_e \rightarrow \nu_{\mu'}$ conversions. The corresponding resonance condition is given by
\begin{equation}\label{eq:MSWL}
{\rm MSW-L: }~~~~~ \rho_{res} Y_e \simeq {{m_N \over{\sqrt{2} G_F}}{ {\Delta m^2_{21}} \over{2E}} \cos2\theta_{12}} \ . 
\end{equation}

Besides the MSW-H and MSW-L resonances, 
three\footnote{Note that a fourth mechanism was first pointed out in \cite{Akhmedov:2003fu}, referred to as RSFP-X, that depends on the $\cos^2\theta_{13}$. It reduces to the RSFP-E one for small $\theta_{13}$ values.} spin-flavor mechanisms can be present. They are referred to as RSFP-L, RSFP-H and RSFP-E \cite{Akhmedov:2003fu}. 
\begin{figure}
\begin{center}
\includegraphics[scale=0.45]{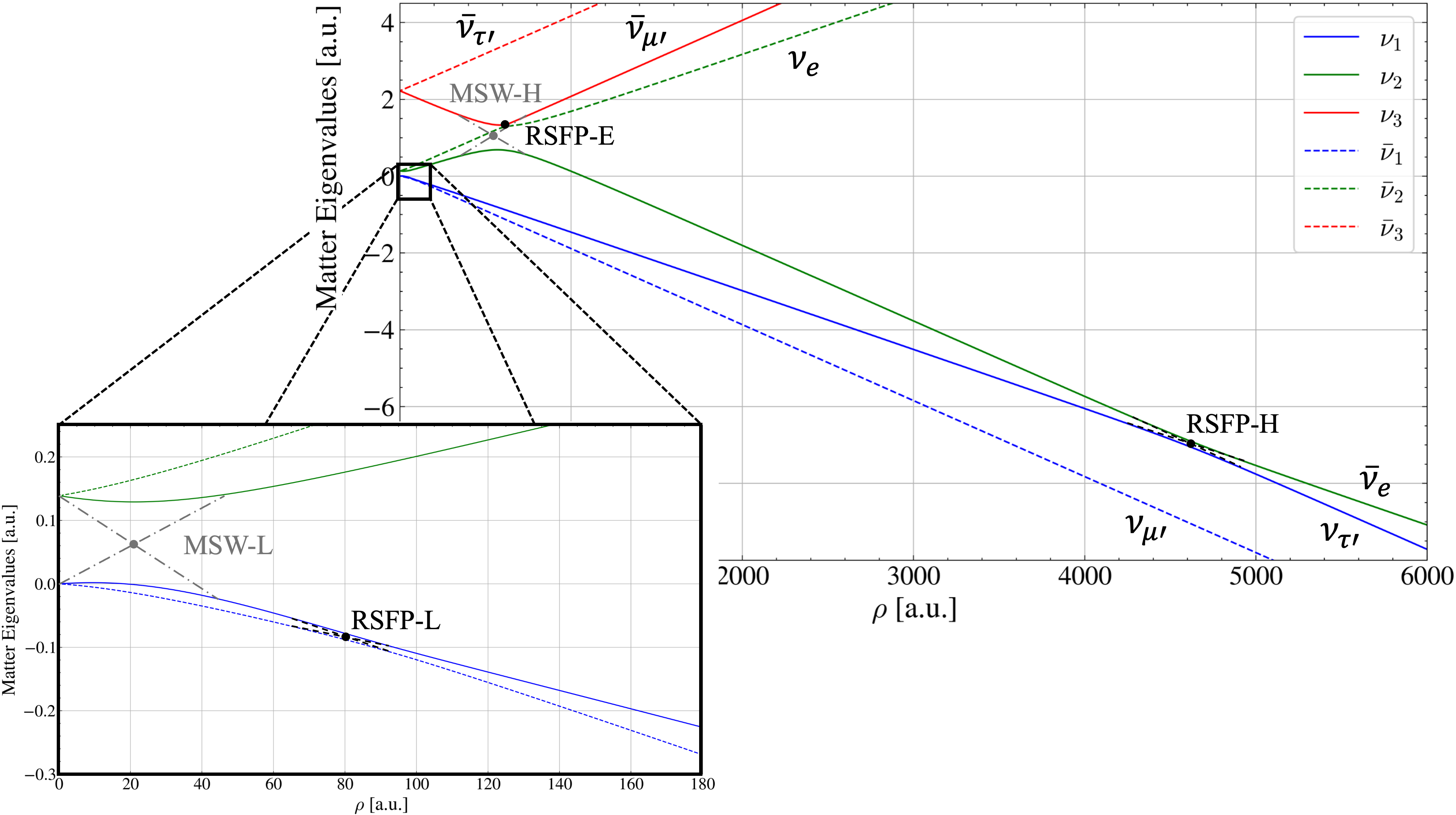} 
\includegraphics[scale=0.45]{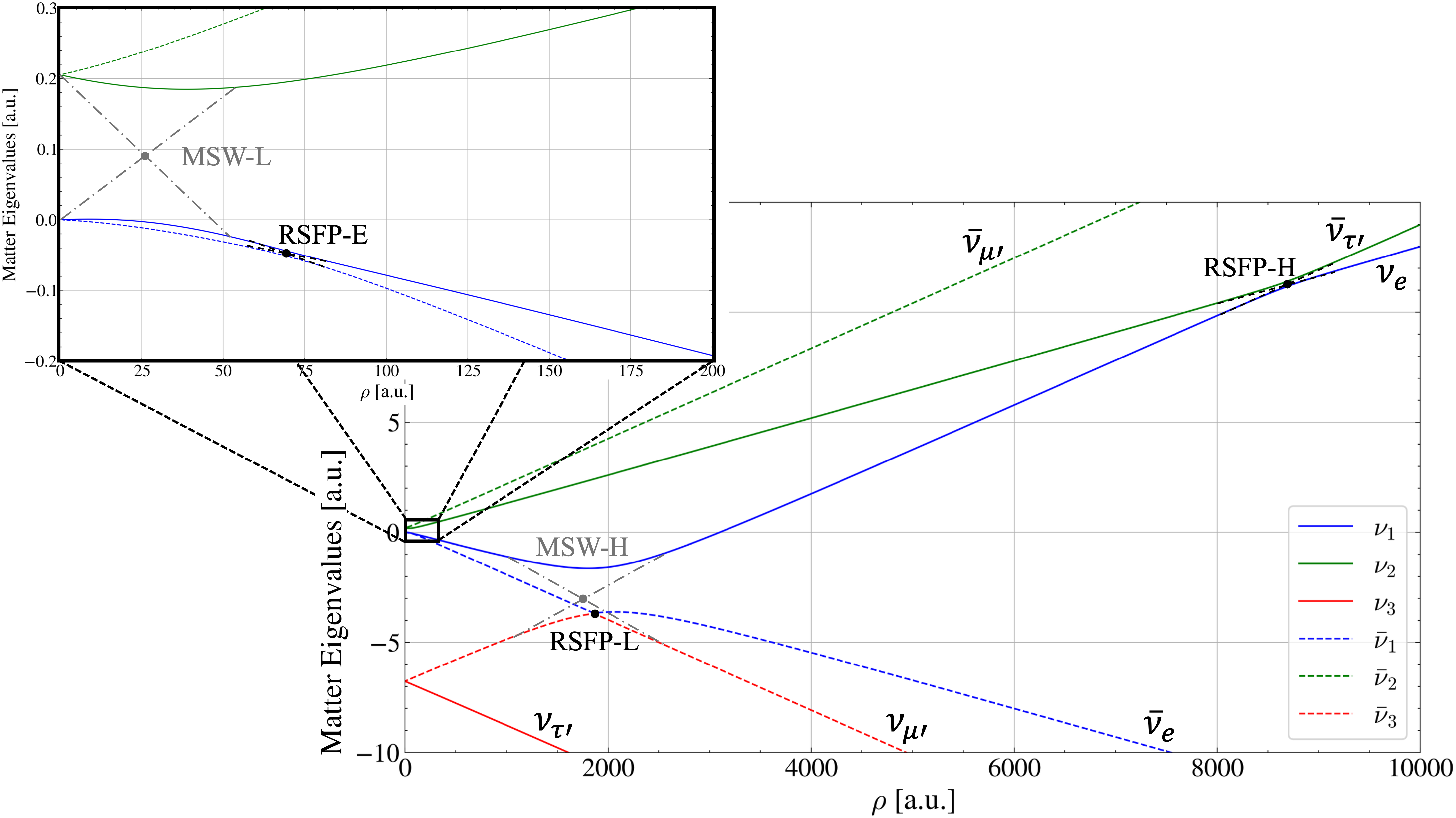} 
\caption{Neutrino matter eigenvalues (in arbitrary units) as a function of the matter density (in arbitrary units), for Majorana neutrinos, in presence of transition magnetic moments coupling to a strong magnetic field. The results are for
the schematic model, where the matter density and the transverse magnetic field profiles are taken as power laws, \eqref{eq:matprof} and \eqref{eq:Bprof} respectively.
The upper figure corresponds to NO, while the lower figure corresponds to IO. The labels show the locations of the high and low MSW resonances (MSW-H and MSW-L) as well as of the resonant spin-flavor precession ones --  RSFP-H, RSFP-L, RSFP-E (see text). The results are for $Y_e = 0.4995$.}
\label{fig:static}
 \end{center} 
\end{figure}
\begin{itemize}
 \item {\bf RSFP-H:} 
The high spin-flavor resonance, associated with $(\theta_{31}, \Delta m^2_{31})$, can produce either $ \bar{\nu}_e  \rightarrow \nu_{\tau}'$
(NO)  or $ \nu_e  \rightarrow \bar{\nu}_{\tau}'$ (IO) conversions. The RSFP-H resonance condition is
\begin{equation}\label{eq:RSFP-H}
 \rho_{res} (1 - 2 Y_e) \simeq {{m_N} \over {\sqrt{2} {G_F}}}  {{\Delta m^2_{31}} \over {2E}} \cos2\theta_{13} \ . 
\end{equation}

\item {\bf  RSFP-L:} 
The low spin-flavor resonance, associated with $(\theta_{21}, \Delta m^2_{21})$, produces $\bar{\nu}_e \rightarrow \nu_{\mu'}$, conversions both for NO and for IO. The RSFP-L condition reads
\begin{equation}\label{eq:RSFP-L}
 \rho_{res} (1 - 2 Y_e) \simeq {{m_N} \over {\sqrt{2} {G_F}}}  {{\Delta m^2_{21}} \over {2E}} \cos2\theta_{12} \ .
\end{equation}

\item {\bf  RSFP-E: }
This spin-flavor resonance, first identified in ref.~\cite{Akhmedov:2003fu}, arises from a pure 3$\nu$ flavor effect, so that its location depends on both $(\theta_{21}, \Delta m^2_{21})$ and $(\theta_{31}, \Delta m^2_{31})$. The RSFP-E induces $ \nu_e  \rightarrow \bar{\nu}_{\tau}'$ (NO) or $ \bar{\nu}_e  \rightarrow \nu_{\tau}'$ (IO) conversions. From our numerical calculations, it turns out that the approximate expression given in  \cite{Akhmedov:2003fu} does not appear accurate enough to identify analytically the RSFP-E location.  
\end{itemize}

Figure \ref{fig:static} displays the numerical results for the neutrino matter eigenvalues as a function of the matter density. We assume for the electron fraction the value $Y_e$ = 0.4995.
One can see the RSFP-H resonance 
is present both for NO and IO.
As for the RSFP-L resonance, it appears close to the MSW-L resonance in NO, whereas it is close to the MSW-H resonance for IO. For the RSFP-E resonance the reverse it true.

As known, once a resonance condition is fulfilled, 
the neutrino flavor transformation at the resonance depends upon its width and upon the adiabaticity of the neutrino evolution through the resonances. It is to be noted that to evaluate the impact of neutrino flavor transformation at the resonances
some of the previous works, e.g. \cite{Akhmedov:2003fu,Jana:2022tsa}, used the RSFP adiabaticity parameters only, that for RSFP-H and RSFP-L reads
\begin{equation}\label{eq:adspin}
\gamma^{\rm RSFP-H(L)}_{res} = {{8 r_0} \over 3} \Big({{E} \over {|\Delta m^2|}} \Big)^{4/3} \Big[{{2 \sqrt{2} G_F \rho_0  |1 - 2 Y_e|}  \over {m_N \cos2\theta}} \Big]^{1/3} (\mu {\rm B}_{\perp, res})^2 \ , 
\end{equation}
where ${\rm B}_{\perp,res}$ is the magnetic field at the resonance. Thus
the adiabaticity depends strongly on the $\mu {\rm B}$ value at the resonance. We remind that the MSW adiabaticity parameter reads
\begin{equation}
\gamma^{\rm MSW}_{res} = {r_0 \over 6} \Big({ \Delta m^2 \over E}\Big)^{2/3} {\sin^2 2\theta \over (\cos 2 \theta)^{4/3}}\Big[{{2 \sqrt{2} G_F \rho_0 Y_e}  \over {m_N}} \Big]^{1/3} \ . 
\end{equation}

\begin{figure}
\begin{center}
\includegraphics[scale=0.5]{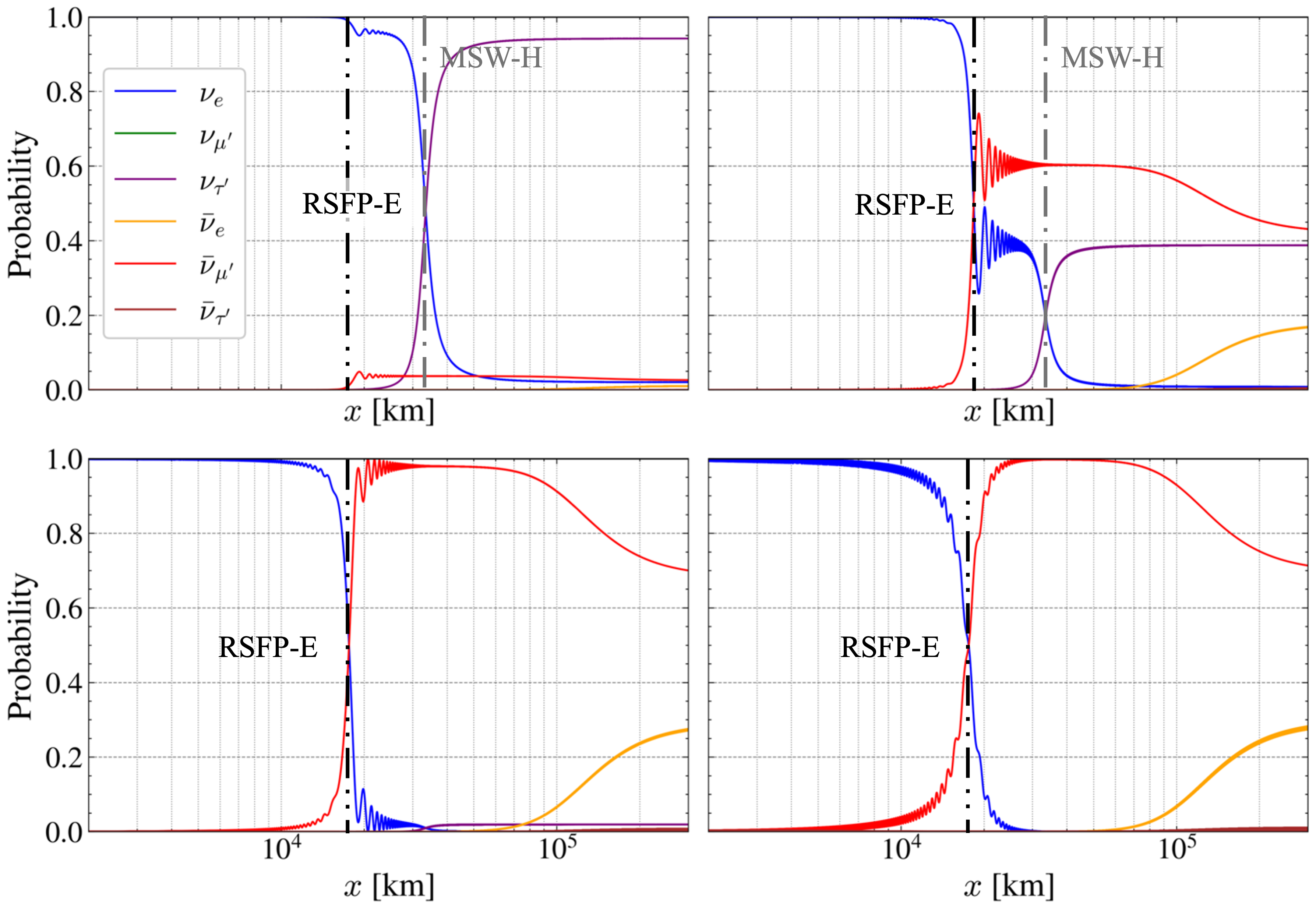} 
 \end{center}
 \caption{Neutrino and antineutrino probabilities, as a function of distance in a supernova, including non-zero transition magnetic moments for Majorana neutrinos. The results are obtained in the schematic model, in which the matter density and the transverse magnetic field profiles are taken as power laws, \eqref{eq:matprof} and \eqref{eq:Bprof} respectively. The case 
 NO is shown. The different figures correspond to the following cases:  
 {\it i)} $ \mu {\rm B}_{0} = 10~ \mu_B {\rm G}$ (upper left);  {\it ii)} $ \mu {\rm B}_{0} = 50~ \mu_B {\rm G}$ (upper right);  {\it iii) } $ \mu {\rm B}_{0} = 100~ \mu_B {\rm G}$ (lower left);  {\it iv)} $ \mu {\rm B}_{0} = 200~ \mu_B {\rm G}$ (lower right figure).}
 \label{fig:proba_NO}
\end{figure}

\begin{figure}
\begin{center}
\includegraphics[scale=0.5]{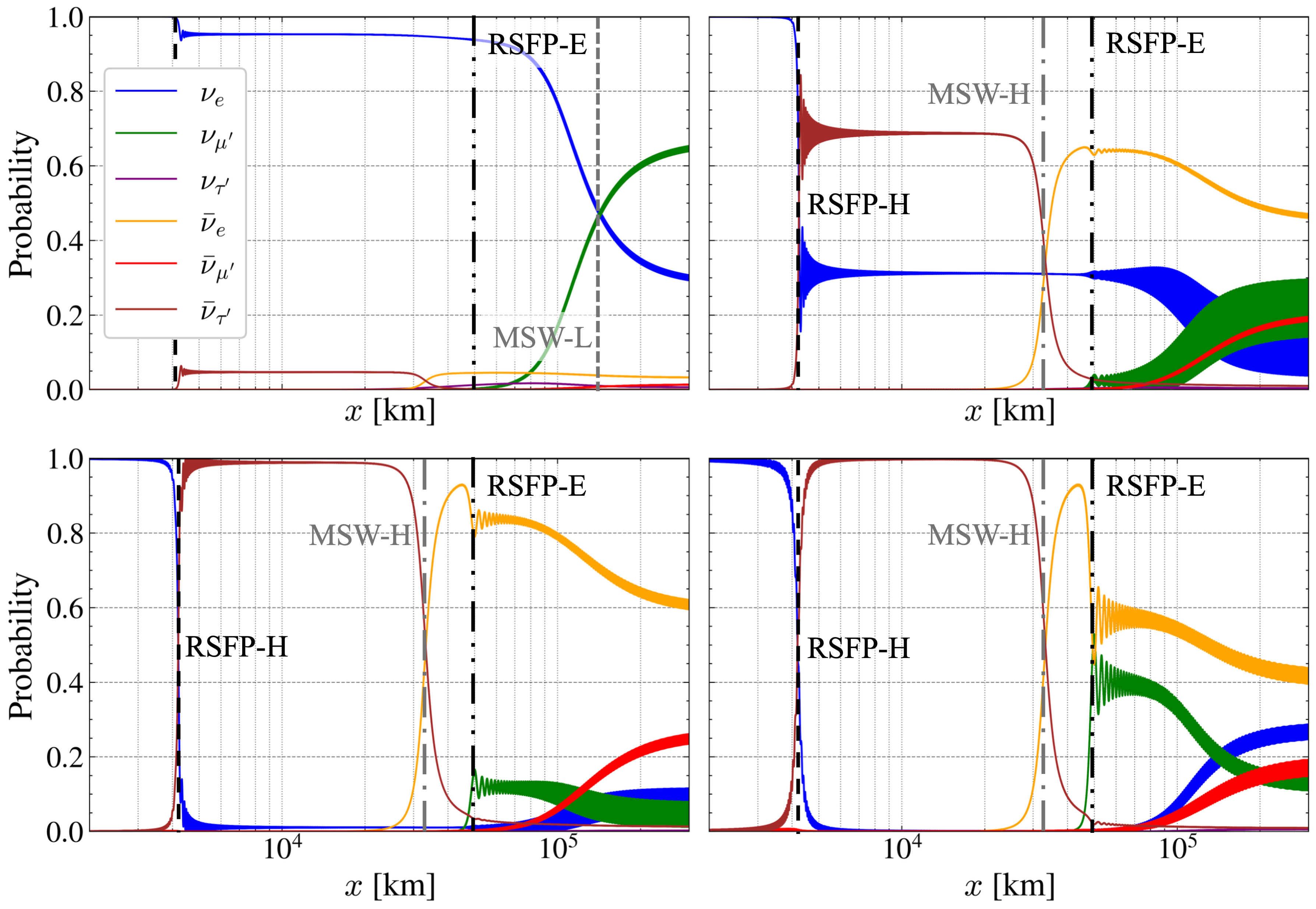} 
 \end{center}
 \caption{Neutrino and antineutrino probabilities, as a function of distance in a supernova in presence of non-zero transition magnetic moments for Majorana neutrinos. The results are for the case of the schematic model, in which the matter density and the magnetic field strength profiles  are taken as power laws, \eqref{eq:matprof} and \eqref{eq:Bprof} respectively. The case 
 IO is shown. The figures correspond to the following cases:  
 {\it i)} $ \mu {\rm B}_{0} = 10~ \mu_B {\rm G}$ (upper left);  {\it ii)} $ \mu {\rm B}_{0} = 50~ \mu_B {\rm G}$ (upper right);  {\it iii) } $ \mu {\rm B}_{0} = 100~ \mu_B {\rm G}$ (lower left);  {\it iv)} $ \mu {\rm B}_{0} = 500~ \mu_B {\rm G}$ (lower right figure). }
 \label{fig:proba_IO}
\end{figure}

The neutrino probabilities are shown in figures \ref{fig:proba_NO} and \ref{fig:proba_IO} for NO and IO respectively,
and different values of the  $ \mu {\rm B}_{0}$ products.
The $ \mu {\rm B}_{0}$ range of values is taken to cover both adiabatic and non-adiabatic evolution at the resonances.

In the NO case, the RSFP-E impact is visible with a neutrino evolution at resonance that is ranging from non-adiabatic ($ \mu {\rm B}_{0} = 10~\mu_B {\rm G}$) to fully adiabatic for $ \mu {\rm B}_{0} >  50~\mu_B {\rm G}$. Evolution at the MSW-H resonance is clearly visible and important in particular when the RSFP-E is not very large. 

As for the IO case, from figure \ref{fig:proba_IO}, one can see that efficient $\nu_e \rightarrow \nu_{\mu'}$ conversions due to the MSW-L resonance are visible in all plots. This is also the case for the $\bar{\nu}_e \rightarrow \bar{\nu}_{\tau'}$ conversions due to the MSW-H resonance. As for the spin-flavor resonances, the $\nu_e \rightarrow \bar{\nu}_{\tau'}$ (semi-) adiabatic conversions due to the RSFP-H resonance are clearly visible when $ \mu {\rm B}_{0} \geq~ 50~\mu_B {\rm G}$; whereas RSFP-E adiabatic (semi-adiabatic) conversions between $\bar{\nu}_e$ and $\nu_{\mu'}$ can be seen for 
$ \mu {\rm B}_{0} \geq ~ 500 ~(100) ~\mu_B {\rm G}$.

\subsection{Impact of the magnetic moment on the neutrino fluxes}
At the neutrinospheres, the neutrino differential number fluxes for each neutrino flavor $\nu_{\alpha}$ are given by 
\begin{equation}\label{eq:flux}
\phi^0_{\nu_{\alpha}}(E)
=
\frac{L_{\nu_{\alpha}}}
{\langle E \rangle_{\nu_{\alpha}}}
f_{\nu_{\alpha}}(E) \ ,
\end{equation}
where $L_{\nu_{\alpha}}$ is the neutrino luminosity and
$\langle E \rangle_{\nu_{\alpha}}$ the neutrino mean energy. The neutrino energy distribution $f_{\nu_{\alpha}}(E)$ is given by a quasi-thermal spectrum. It is well accounted for by the "power law" distribution \cite{Keil:2002in}
\begin{equation}\label{eq:pwflux_new}
f_{\nu_{\alpha}}(E)
=
\frac{(\alpha'+1)^{\alpha'+1}}  {\Gamma(\alpha'+1)\,\langle E\rangle_{\nu_{\alpha}}}
\left(
\frac{E}{\langle E\rangle_{\nu_{\alpha}}}
\right)^{\alpha'}
\exp\left[
-\frac{(\alpha'+1)E}{\langle E \rangle_{\nu_{\alpha}}}
\right] \ , 
\end{equation}
with the pinching parameter $\alpha'$ 
\begin{equation}\label{eq:pinch}
\alpha' = {{\langle E^2 \rangle_{\nu_{\alpha}} - 2 \langle E \rangle_{\nu_{\alpha}}^2} \over {\langle E \rangle}_{\nu_{\alpha}}^2 - \langle E^2 \rangle_{\nu_{\alpha}}} \ , 
\end{equation}
The neutrino spectrum satisfies the normalization condition
\begin{equation}\label{eq:pwfluxnor}
\int_0^{\infty} f_{\nu_{\alpha}}(E) dE = 1 \ .
\end{equation}

By solving \eqref{eq:eveq3nu}, we determine the evolution through the stellar layers for 3$\nu$ flavors. Table \ref{tab:fluxpar} presents the parameters used for the neutrino fluxes - luminosities and mean energies. 
On Earth
the neutrino fluxes are given by
\begin{equation}\label{eq:fluxfi}
\phi_{\nu_{\alpha}}(E) = {1 \over {4 \pi d^2}} \sum_i \vert U_{\alpha i} \vert^2 \phi_{\nu_i} (E) \ , 
\end{equation}
where $d$ is the core-collapse supernova distance
that we take equal to 10 kpc. 

\begin{table}\label{tab:flux} 
    \centering
\begin{tabular}{lcc}
\toprule
 & Luminosity & Mean energy \\
\midrule
$\nu_e$ & 2.14 &  12.06 \\
$\bar{\nu}_e $ & 2.25 & 16.11 \\
$\bar{\nu}_x $& 1.68 & 21.16 \\
\bottomrule
\end{tabular}
\caption{Schematic model: Neutrino luminosities (in units of $10^{52}~{\rm erg} \cdot {\rm s}^{-1}$) and mean energies (in MeV) defining the neutrino fluxes \eqref{eq:flux}-\eqref{eq:pinch}.}\label{tab:fluxpar}
\end{table}

\begin{figure}
  \centering
  \begin{subfigure}[b]{0.48\textwidth}
    \centering
    \includegraphics[width=\linewidth]{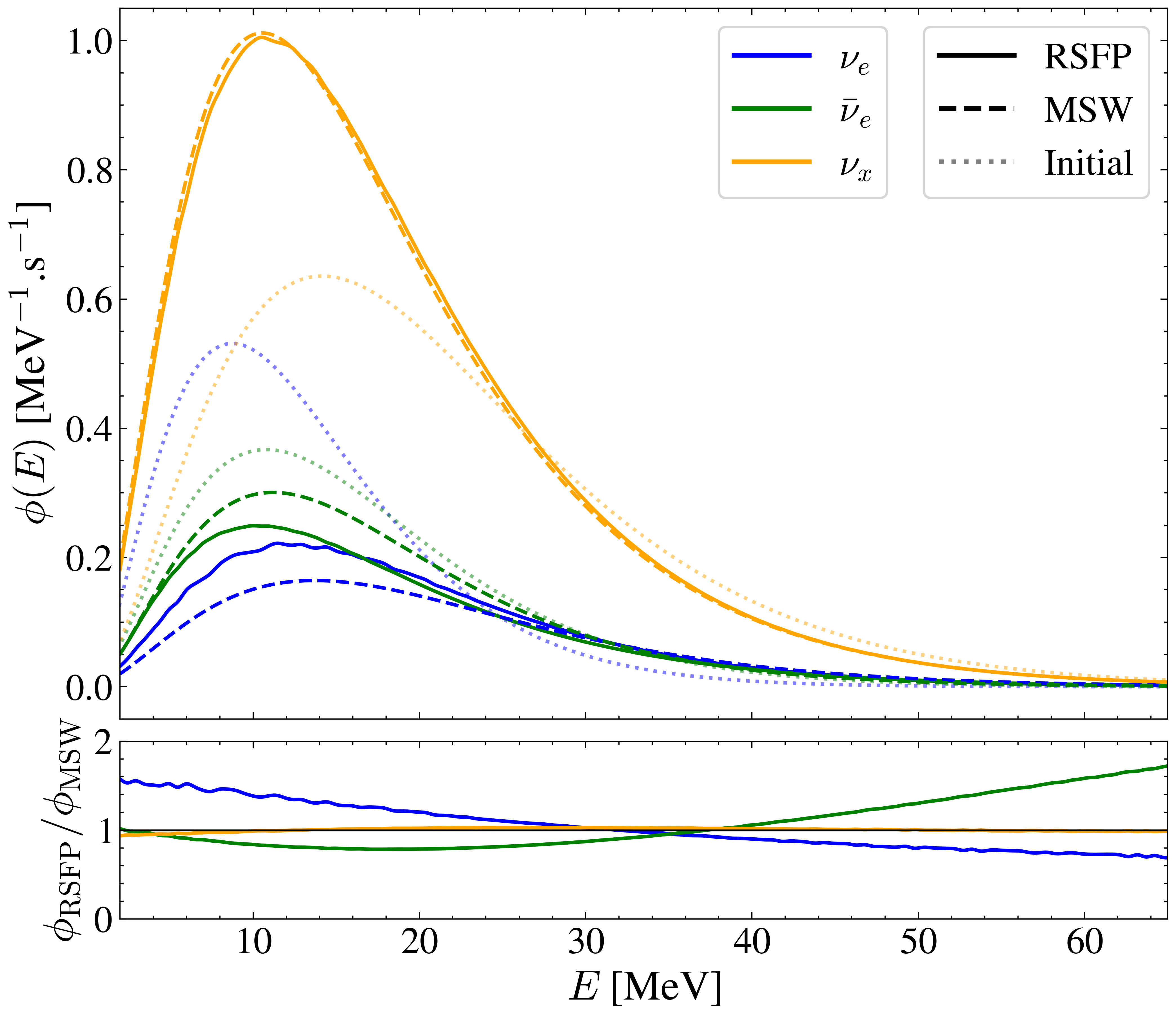}
    \caption{NO}
    \label{fig:sub1}
  \end{subfigure}
  \hfill
  \begin{subfigure}[b]{0.48\textwidth}
    \centering
    \includegraphics[width=\linewidth]{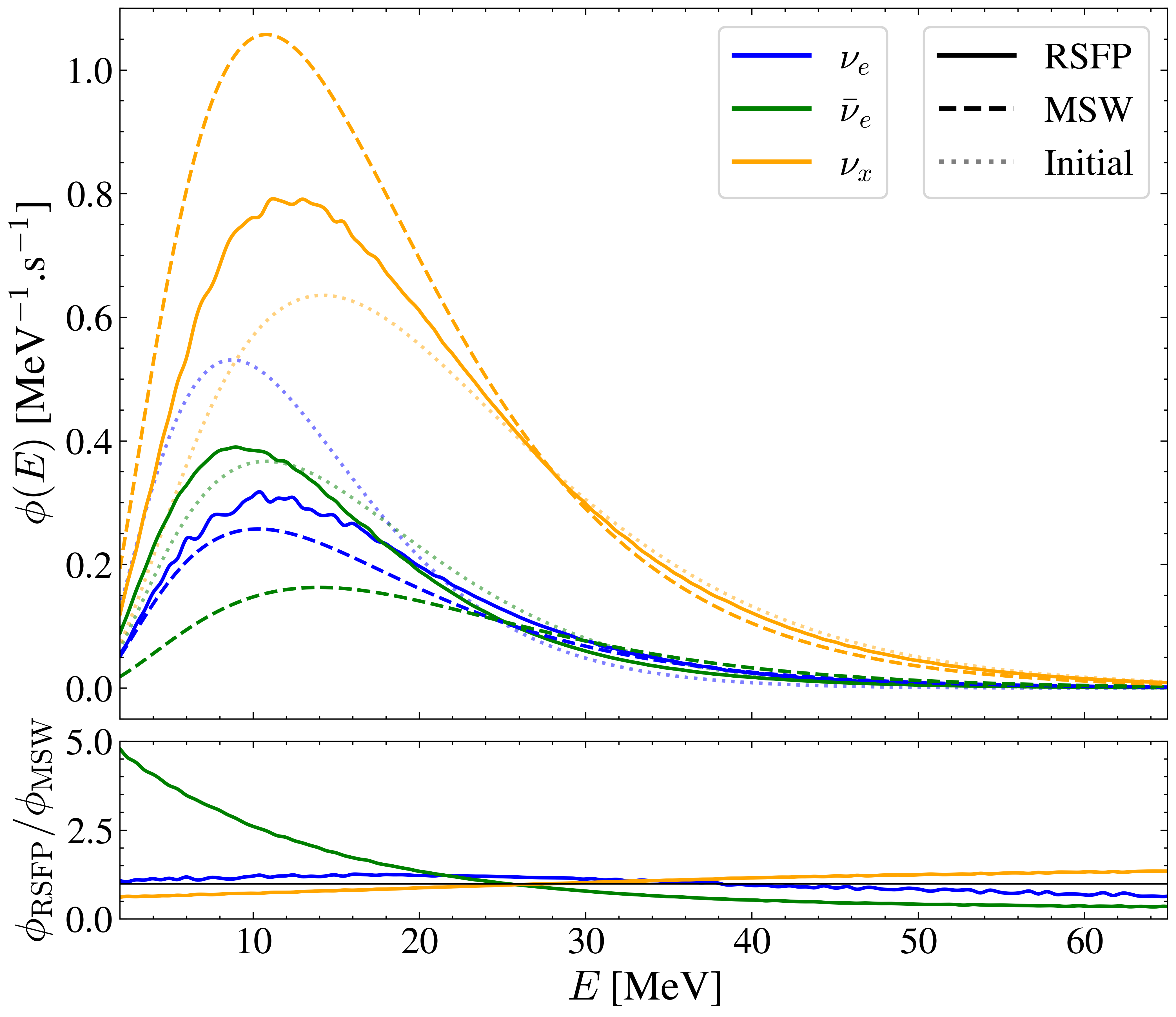}
    \caption{IO}
    \label{fig:sub2}
  \end{subfigure}
  \caption{Neutrino fluxes on Earth for the different neutrino species, after propagating through the MSW and RSFP resonances, in a core-collapse supernova. Note that the non-electron neutrino flux noted as $\nu_x$ here is the summed flux over all non-electron flavors.  The results correspond to the schematic model for NO and $\mu {\rm B}_0 =  200~\mu_B {\rm G}$ (left figure) and for IO and $\mu {\rm B}_0 = 100  ~\mu_B {\rm G}$ (right figure). For comparison the initial fluxes, before flavor evolution \eqref{eq:flux}-\eqref{eq:pinch}, are also shown. The lower figures present the ratio between the fluxes in presence of RSFP and the ones with the MSW effect only.}
  \label{fig:globale}
\end{figure}

Figure \ref{fig:globale} presents the impact of neutrino-antineutrino mixings on the $\nu_e, \bar{\nu}_e$ and $\nu_x$ fluxes. The cases shown are NO and IO, for 
$\mu {\rm B}_0 = 200 ~\mu_B {\rm G}$ and $\mu {\rm B}_0 = 100 ~\mu_B {\rm G}$ respectively. One can see that the inclusion of transition magnetic dipole moments modifies the spectra by a factor of 2 both at low and high neutrino energies for NO; whereas for IO, and $E < 20 $ MeV, the effect is larger than a factor of 4. This is mostly due to the RSFP-E resonance. 
Similar quantitative behaviors are obtained for 
$\mu {\rm B}_0 = 100 ~\mu_B {\rm G}$ in the NO case and $\mu {\rm B}_0 = 200 ~\mu_B {\rm G}$ in the IO one. 

Before applying the spin-flavor framework discussed above to neutrino and magnetic-field profiles in a more realistic case, we introduce now the
multidimensional core-collapse simulations used in the present study.

\section{Three-dimensional supernova models}
\label{Sect:simulations}

Two one-dimensional stellar-evolution progenitors from \cite{AguileraDena:2018} (models A17 and A26 hereafter) were  mapped to three dimensions and their collapse was followed with the neutrino-radiation magnetohydrodynamics (MHD) code \textsc{Aenus--Alcar}, using the same numerical framework as in refs.~\cite{Obergaulinger:2021omt,ObergaulingerAloy:2021PaperIII}. \textsc{Aenus--Alcar} evolves the special-relativistic ideal-MHD equations with a finite-volume, high-resolution shock-capturing scheme, and solves the energy-dependent neutrino-transport equations. General-relativistic effects on the hydrodynamics are included approximately through the effective Tolman--Oppenheimer--Volkoff potential (version A of \cite{Marek:2005if}).  The high-density matter is described by the SFHo equation of state \cite{Steiner:2012rk}; at low density, an equation of state for photons, electrons, positrons, and nucleons is combined with a nuclear-statistical-equilibrium treatment or a nuclear-flashing prescription, as appropriate \cite{Obergaulinger:2021omt}.

The neutrino transport is formulated as a spectral two-moment scheme: evolution equations for the neutrino energy and momentum densities are closed with a local algebraic M1 closure for the pressure tensor (see \cite{Just:2015fda} for details of the implementation).  Separate moment systems are evolved for electron neutrinos, electron antineutrinos, and a representative non-electron species, $\nu_x$, standing for muon and tau neutrinos and antineutrinos.  Velocity-dependent and gravitational terms couple the energy groups consistently to order $\mathcal{O}(v/c+)$ \cite{Endeve:2012dt}.  The interaction set comprises emission, absorption, and scattering on free nucleons and nuclei, inelastic scattering off electrons, electron--positron pair annihilation, and nucleon--nucleon bremsstrahlung \cite{Obergaulinger:2021omt}.

The reference three-dimensional setup employs $(n_r,n_\theta,n_\phi)=(300,64,128)$ zones in spherical polar coordinates. The radial grid is logarithmically spaced, and the MHD  variables and neutrino moments are reconstructed with a fifth-order monotonicity-preserving method.  The neutrino spectrum is discretised into $n_\epsilon=10$ (model A17) or $n_\epsilon=6$ (model A26) logarithmically spaced energy bins covering $3\mbox{--}240\,{\rm MeV}$ \cite{ObergaulingerAloy:2021PaperIII}.  

The remainder of this section is organised as follows.  We first introduce the
stellar progenitors (section~\ref{sec:stellar-progenitors}) and then consider two complementary explosion scenarios:
model A17, in which shock revival is primarily neutrino-driven and is followed
by magnetically powered proto-magnetar activity (section~\ref{sec:A17}), and model A26, whose explosion
is predominantly magnetorotational, although neutrino heating also contributes
to shock revival (section~\ref{sec:A26}). We note that the stellar evolution models have not explicitly been evolved in 3D prior to collapse as in refs.~\cite{Griffiths_I_2026arXiv260522927,Griffiths_II_2026arXiv260522938}. Thus, the magnetic field topology and strength are as provided by the one-dimensional secular evolution.

\subsection{Stellar progenitor models}
\label{sec:stellar-progenitors}

Models A17 and A26 belong to the rapidly rotating, low-metallicity Series~B progenitors of \cite{AguileraDena:2018}. The models start on the zero-age main sequence with masses of $17\,\mathrm{M}_{\odot}$ and $26\,\mathrm{M}_{\odot}$, respectively, an equatorial velocity of $600\,\mathrm{km\,s^{-1}}$, and a metallicity $Z=Z_{\odot}/50$.  Their rapid rotation leads to quasi-chemically homogeneous evolution.  Rotational mixing and rotationally enhanced mass loss remove the hydrogen envelope and leave only about $10^{-2}\,\mathrm{M}_{\odot}$ of helium, so that both progenitors are compact, nearly bare carbon--oxygen stars (with less than $0.01\,M_\odot$ of helium in their envelopes) and natural candidates for Type~Ic engine-driven transients. 

The principal precollapse properties are summarised in Table~\ref{tab:bonn-progenitors}. We quote the values from the updated calculations of \cite{AguileraDena:2020}; these calculations use the same physical scenario as the 2018 Series~B models, but employ an improved treatment of angular-momentum loss and reach the onset of collapse. 

\begin{table}[t]
\centering
\begin{tabular}{lcc}
\toprule
Property & A17 & A26 \\
\midrule
$M_{\rm ZAMS}\,[\mathrm{M}_{\odot}]$ & 17 & 26 \\
$M_{\rm final}\,[\mathrm{M}_{\odot}]$ & 13.39 & 20.05 \\
$\bar{j}_{1.5}\,[10^{15}\,\mathrm{cm^2\,s^{-1}}]$ & 1.33 & 1.75 \\
$\bar{j}_{2}\,[10^{15}\,\mathrm{cm^2\,s^{-1}}]$ & 2.58 & 3.19 \\
$\bar{j}_{5}\,[10^{15}\,\mathrm{cm^2\,s^{-1}}]$ & 8.96 & 9.29 \\
$\xi_{2.5}$ & 0.59 & 0.36 \\
\bottomrule
\end{tabular}
\caption{Selected precollapse properties of the A17 and A26 progenitors from \cite{AguileraDena:2020}. $M_{\rm ZAMS}$ is the initial stellar mass at the zero-age main sequence, whereas $M_{\rm final}$ is the total stellar mass at the onset of core collapse. 
Mass-averaged angular momenta contained within the Lagrangian mass coordinates $1.5\,M_{\odot}$, $2\,M_{\odot}$, and $5\,M_{\odot}$, are respectively denoted by $\bar{j}_{1.5}$, $\bar{j}_{2}$, and $\bar{j}_{5}$. Finally, $\xi_{2.5}$ is the dimensionless precollapse compactness parameter evaluated at the Lagrangian mass coordinate $2.5\,M_{\odot}$. A larger value of
$\xi_{2.5}$ therefore corresponds to a more compact inner stellar core.}
\label{tab:bonn-progenitors}
\end{table}

\subsection{Neutrino-driven supernova model A17}
\label{sec:A17}

Model A17 develops a neutrino-driven explosion.  As the shock expands, accretion onto the PNS star subsides, its mass and rotational energy at a post-bounce time of $t_{\mathrm{pb}} = 3.1 \, \mathrm{s}$ are $M_{\mathrm{PNS}} = 1.94 \, M_{\odot}$ and $E_{\rm PNS}^{\rm rot} =  1.1 \times 10^{51}\, \mathrm{erg}$, respectively.  The remnant remains a PNS throughout the simulated evolution.  An explosion sets in at about $t_{\mathrm{pb}} \approx 0.5 ~ \mathrm{s}$.  It is driven predominantly by neutrino heating.  In the following about $1 \, \mathrm{s}$, the explosion energy increases to $E_{\mathrm{exp}} \approx 3 \times 10^{50} \, \mathrm{erg}$.  Afterwards, the growth of $E_{\mathrm{exp}}$ slows down.  By the end of the simulation, it reaches $E_{\mathrm{exp}} \approx 4.5 \times 10^{50} \, \mathrm{erg}$.  A large part of this late rise comes from the gradual extraction of rotational energy of the PNS by its magnetic field and is carried by wide outflows along the rotational axis.

The PNS develops and sustains convection in a region between radii of about $10 - 20 \, \mathrm{km}$.  Its rotation is strongly differential with a maximum angular velocity $\Omega$ at the lower border of the convective layer or a few $\mathrm{km}$ below that and a negative gradient $\partial_{r} \Omega < 0$ extending from there to the PNS surface.
The magnetic energy inside the PNS is dominated by the toroidal component with maximum values of ${\rm B}_{\mathrm{tor,max}} \sim 10^{15} \, \mathrm{G}$, which is typically about half an order of magnitude stronger than the poloidal component.  
The toroidal component is strongest where the rotational shear is most pronounced, i.e., near the bottom of the convective zone and closer to the centre.  The former of these two regions is also where we find the strongest poloidal field.

\subsection{Magnetorotational supernova model A26}
\label{sec:A26}

Model A26 also develops a delayed explosion.  Slow shock expansion sets in after $t_{\mathrm{pb}} \gtrsim 0.1 \, \mathrm{s}$ and speeds up about $0.5 \, \mathrm{s}$ afterwards. Over a period of $\approx 0.3 \, \mathrm{s}$, the explosion energy rises from $E_{\mathrm{exp}} = 10^{50} \, \mathrm{erg}$ to $E_{\mathrm{exp}} = 5 \times 10^{50} \, \mathrm{erg}$.  Compared to model A17, the faster growth of $E_{\mathrm{exp}}$ reflects the larger contribution of rotation and magnetic fields to powering the explosion during this phase.  During the subsequent $\approx 0.7 \, \mathrm{s}$, the explosion energy continues to rise, albeit slightly slower, to a final value of $E_{\mathrm{exp}} \approx 9 \times 10^{50} \, \mathrm{erg}$.  The magnetorotational stresses driving the explosion give the ejecta a bipolar geometry characterised by jets which propagate along the rotational axis.  While they roughly retain their propagation direction, they are not strong enough to pierce through the surrounding matter in a straight line and instead widen to large lobes after a distance of $\gtrsim 5000 \, \mathrm{km}$.
In contrast to A17, continued accretion drives the PNS to substantially higher masses, eventually above the limit for gravitational stability.  It reaches a maximum of $M_{\mathrm{PNS}} \approx 2.75\,M_\odot$ after only $\simeq 1.3\,\mathrm{s}$ of evolution post-bounce and then collapses to a black-hole, at which point we had to terminate the simulation.  

The model exhibits the fastest rotation, reaching a final $E_{\rm PNS}^{\rm rot} \approx 5 \times 10^{52}\,$erg. This value is in the empirical threshold identified by \cite{Aloy:2020svw}, which separates models that retain a rapidly rotating PNS from those that undergo collapse to a black-hole. The large attained rotational energy causes the PNS to flatten and acquire an equatorial bulge supported to a large part centrifugally.   PNS convection develops similarly to model A17, but remains slightly slower while covering a larger radial extent, in particular during the late phases in which the PNS gets gradually more compact.  Consequently, we do not find the same marked imprint of the convection zone on the rotational profile as in the previous model.  The angular velocity is distributed more evenly across the interior of the PNS with a very broad maximum between $10$ and $30 \, \mathrm{km}$.  
The magnetic field is dominated by the toroidal component whose average strength, again, exceeds that of the poloidal component by about half an order of magnitude.  Both components assume their maximum strength inside the PNS at radii $r \approx 10 \, \mathrm{km}$.  We find that they extend significantly beyond the PNS surface, which is what enables them to extract rotational energy from the PNS and power the explosion.

\section{Impact of neutrino-antineutrino conversions: numerical results}
\label{Sect:results}

Having presented aspects of the multidimensional simulations used, we now go back to our 
investigation on the impact of the neutrino magnetic moment for Majorana neutrinos. 
Let us look at the results for the neutrino fluxes and rates using the neutrino-driven model A17 and the magnetorotational model A26, obtained
by  solving the 3$\nu$ neutrino evolution equations.
The neutrino detectors considered are the running JUNO and the upcoming HK. 
With 258 kt (187 kt fiducial volume), HK, located at at the Tochibora site, will be the largest water Cherenkov detector ever built \cite{Hyper-Kamiokande:2018ofw}. With 8.3 active volume larger than Super-Kamiokande, HK should start taking
data in 2028\footnote{Note that Gadolinium addition is under study.}. 
JUNO is a liquid scintillator detector with 20 kt (18 kt fiducial volume), located in
Jiangmen at 52.5 km from multiple reactor cores \cite{JUNO:2015zny}. The experiment started in August 2025 and already delivered forefront results for the solar oscillation parameters \cite{JUNO:2025gmd} and the first results on $\Delta m^2_{31}$ \cite{Wang}.

The main detection channel in these detectors is inverse beta-decay (IBD) 
\begin{equation}
\bar{\nu}_e + p \rightarrow n + e^+   \ .
\end{equation}
We compute the differential event rate using
\begin{equation}
{{d N_{\bar{\nu}_e}} \over {dt}} = {\epsilon N_t } \int_{E_a}^{E_b} d E' \int_{E_{\rm th(E')}}^{\infty}\phi_{\bar{\nu}_e}(E,t) {{d \sigma_{\rm IBD}} \over {dE'}}(E, E') dE \ ,
\label{eq:event-rate}
\end{equation}
where $\epsilon$ is the detector efficiency, the quantity $N_t$ is the number of targets, $E'=E_{e^+}$ is the positron energy,
$E_{\rm th}(E')$ the energy threshold and
$E_a, E_b$ are the $E_{e^+}$ energy bin bounds.
The positron and neutrino energy are related through $E_{e^+} = E - \Delta_{\rm np}$, with 
$\Delta_{\rm np} = 1.293 ~{\rm MeV}$ the neutron-proton mass difference.
For $\epsilon$, an idealized 100 $\%$ efficiency is taken for the detectors\footnote{Note that results for the running Super-Kamiokande detector can be obtained by scaling the events.}. The number of targets is equal to 
$1.25 \times 10^{34}$ and 
$1.21 \times 10^{33}$ for HK and JUNO respectively. 
The energy threshold is fixed at $E_{\rm th} = 5~ {\rm MeV}$. Finally, we use precise expressions for the IBD cross section from ref.~\cite{Strumia:2003zx} (see also ref.~\cite{Ricciardi:2022pru}). 

\subsection{A17 model: neutrino fluxes and event rates}
\label{sec:A17-results}
First, let us look at the neutrino fluxes and event rate obtained exploiting inputs from the neutrino-driven supernova model A17. Their computation requires, in particular, the matter density and the magnetic field profiles, as well as the parameters defining the neutrino fluxes (luminosities and mean energies\footnote{The pinching parameter $\alpha'$ was kept fixed at a value of 2.}). 

\begin{figure}
\begin{center}
\includegraphics[scale=0.45]{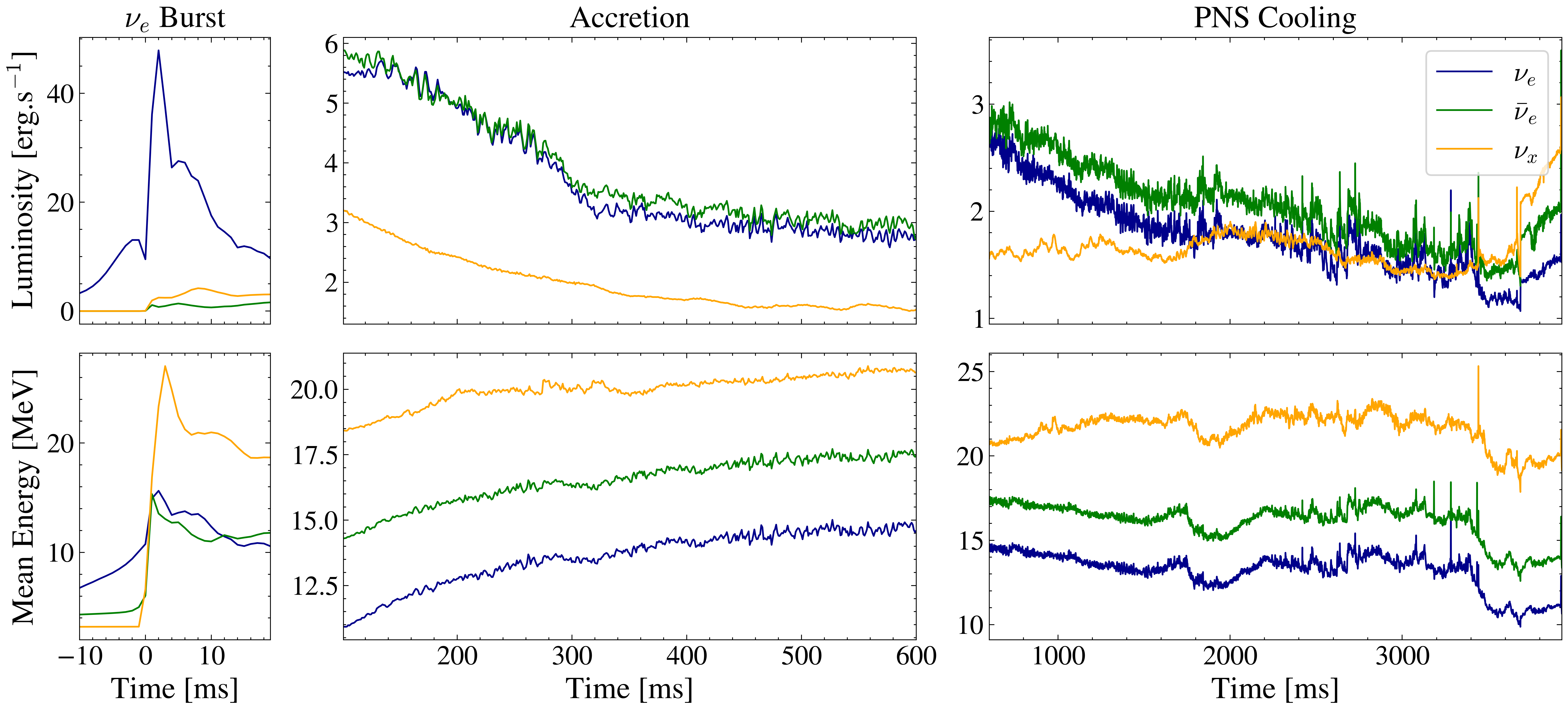} 
 \end{center}
  \caption{Model A17: Neutrino luminosities ($\times 10^{52}$, upper)  
  and mean energies (lower figures) as a function of time for the three-dimensional neutrino-driven supernova model A17, leaving a neutron star. 
  The three figures correspond to the neutronization burst (left),
  accretion phase (middle) and cooling of the PNS star (right). 
The results shown are at $r = 500 ~ {\rm km}$ from the supernova center \cite{Obergaulinger:2021omt,Aloy-Obergaulinger}.}\label{fig:A17}
\end{figure}

Figure \ref{fig:A17} shows the neutrino luminosities and mean energies for the three time windows corresponding to the neutronization burst, the accretion and the cooling of the newly born PNS star. 
One can see that, during the accretion phase and the cooling of the PNS star, the mean energies follow a clear hierarchy with $\langle E_{\nu_e} \rangle < \langle E_{\bar{\nu}_e} \rangle < \langle E_{\nu_x} \rangle$.

\begin{figure}
\begin{center}
\includegraphics[scale=0.5]{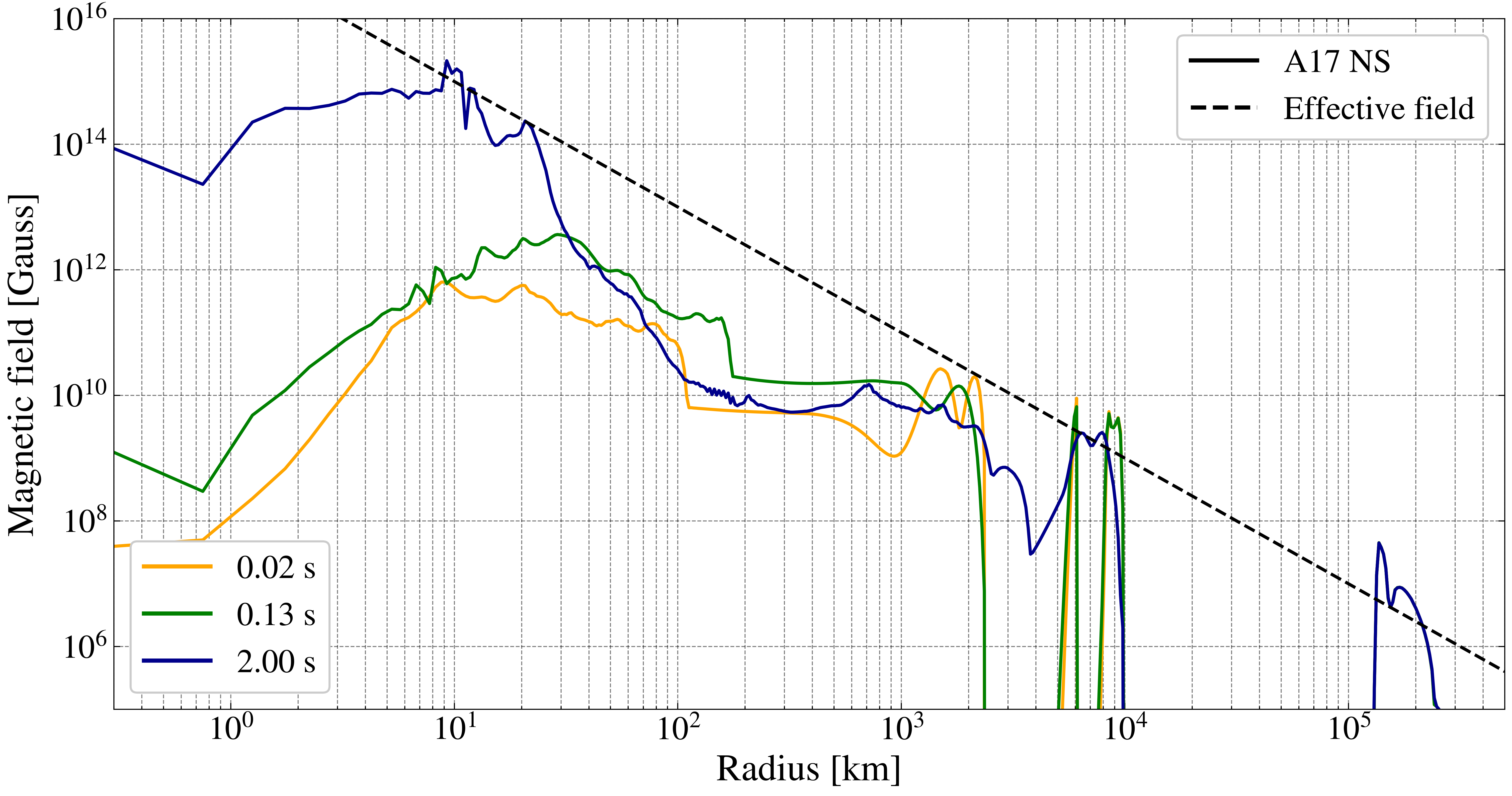} 
 \end{center}
 \caption{Model A17: Angle-averaged transverse magnetic field strength profiles as a function of the distance in the neutrino-driven supernova model A17.
 Three different times after the onset of collapse 
 are shown, namely 0.02 ${\rm s}$ for the neutronization burst, 0.13 ${\rm s}$ for the accretion phase and 2 ${\rm s}$ for the cooling of the newly born PNS star. For comparison, the dashed line indicates the power law magnetic field profile used in the schematic model.}\label{fig:A17mag}
\end{figure}

As for the magnetic field profiles used in our calculations, they are presented, as a function of the radius, in figure \ref{fig:A17mag}, for three different time snapshots. These are chosen as representative for the magnetic field behavior in each time window in figure \ref{fig:A17}. For comparison, we show the power-law profile for the magnetic field. Note that the ${\rm B_0}$ value here is $10^{15}$ Gauss, one order of magnitude smaller than the value employed in the schematic model (section \ref{Sect:schematic}). It is shown "to guide the eye". 
Note that the magnetic field gap in the  $10^{4}$-$10^{5} ~{\rm km}$ range is due to a large convective region in the pre-supernova model, where the stellar-evolution treatment assumes magnetic-flux expulsion, leaving a shell essentially devoid of magnetic field. Whether complete magnetic-flux expulsion occurs under the conditions prevailing during the final stages of stellar evolution remains uncertain. Recent three-dimensional MHD simulations suggest that magnetic fields are amplified in the convective silicon- and oxygen-burning shells, potentially refilling regions predicted to be unmagnetized by one-dimensional stellar-evolution models \cite{Griffiths_II_2026arXiv260522938}. If so, a power law magnetic-field profile used may provide a reasonable approximation to the field strength throughout a substantial fraction of the exploding star.

\begin{figure}
\begin{center}
\includegraphics[scale=0.5]{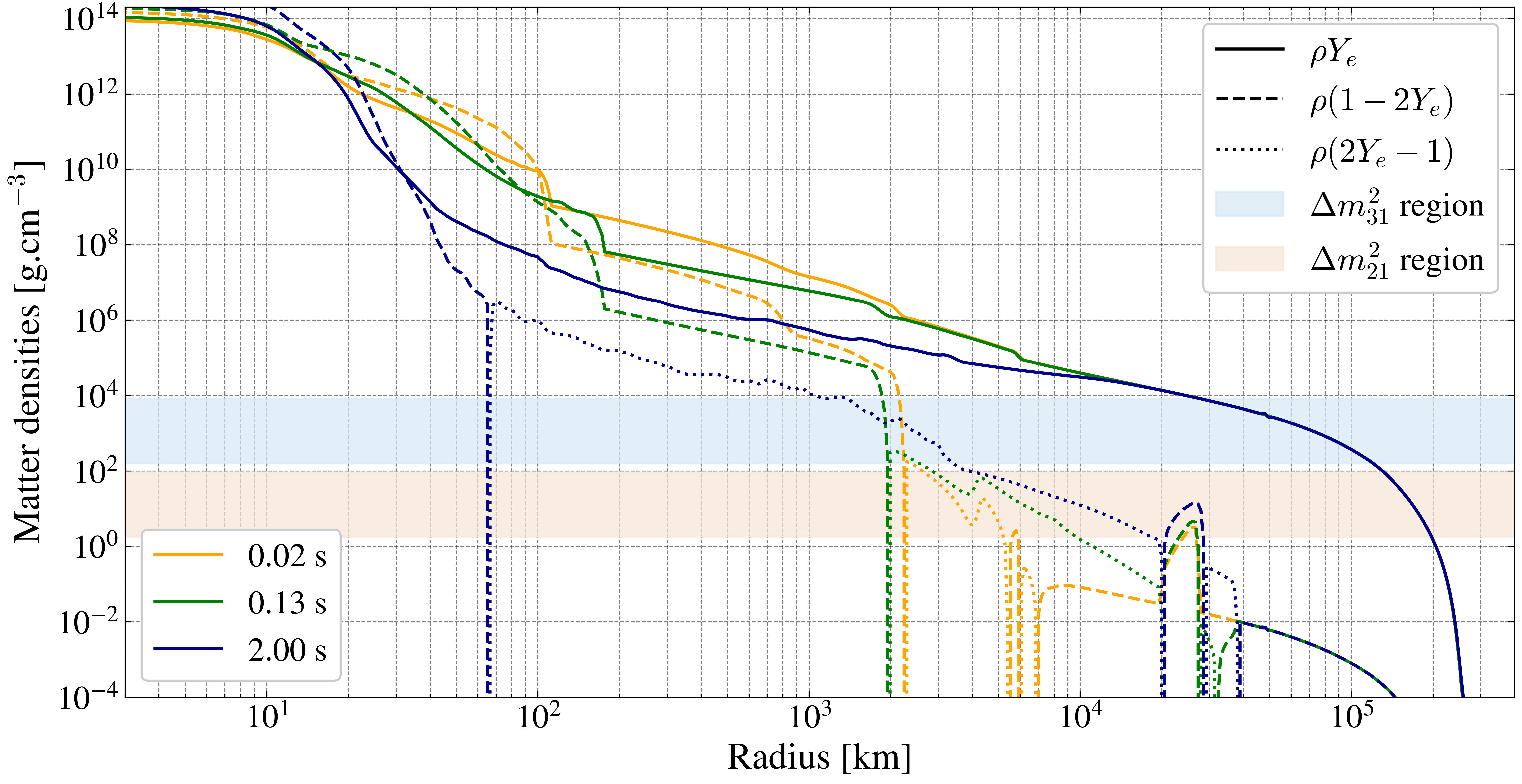} 
 \end{center}
 \caption{Model A17: Angle-averaged matter density profiles as a function of distance, for the three time snapshots of 0.02 s, 0.13 s and 2 s taken as representative for the neutronization burst, the accretion phase and the cooling of the PNS star. The dashed and dotted lines give the {\it l.h.s.} of the RSFP-H \eqref{eq:RSFP-H} and RSFP-L \eqref{eq:RSFP-L} resonance conditions. The colored bands correspond to the {\it r.h.s.} of the same relations as well as the MSW-H \eqref{eq:MSWH} and MSW-L \eqref{eq:MSWL} ones for a typical range of energies $E \in [2, 100]$ MeV. 
 }\label{fig:A17res}
\end{figure}

In order to show the (approximate) locations\footnote{The resonance locations should only be considered as approximate, since the relations \eqref{eq:MSWH}-\eqref{eq:RSFP-L} are strictly valid only for the $2\nu$ case. Corrections for 3$\nu$ flavors are small.} of the spin-flavor resonances, we present the matter density profiles $\rho Y_e$ as a function of the radius in figure \ref{fig:A17res}, together with the quantities  
$\pm \rho (2 Y_e -1)$ that enter into the RSFP-H \eqref{eq:RSFP-H} and RSFP-L \eqref{eq:RSFP-L} resonance conditions. The chosen three time snapshots are displayed.
Note that the colored bands in Figure \ref{fig:A17res} correspond to the {\it r.h.s.} of the resonance relations \eqref{eq:MSWH}-\eqref{eq:RSFP-L}, and cover a typical range of neutrino energies, i.e. $E \in [2, 100]$ MeV. 
One can see the some of the RSFP mechanisms might happen located in the region where the ${\rm B}$ gap is present.

From the $\rho Y_e $ radial profile one can see the presence of the shock wave, in particular at 0.02 s and 0.13 s at about 100 km and 200 km, respectively. For the $t = 2~$s profile, the angle-averaging of the matter density profile does not make it visible. Note also that, at this time, the shock wave has not yet reached the outer layers above $10^{4}~$km.    
The quantity $\rho Y_e $ obviously determines the  MSW-L \eqref{eq:MSWL} and the MSW-H \eqref{eq:MSWH} resonance conditions. 

\begin{figure}
\begin{center}
\includegraphics[scale=0.4]{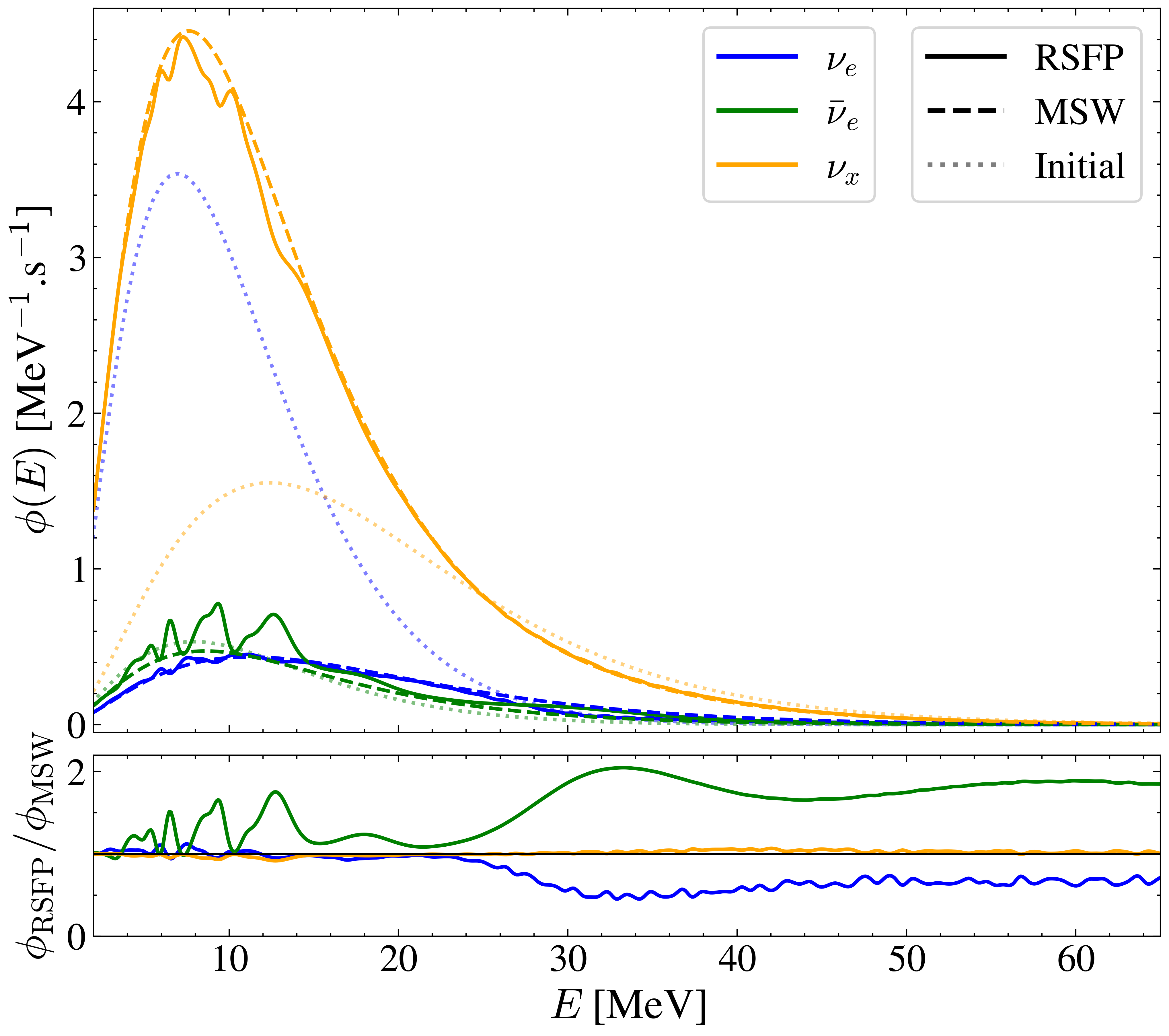} 
\includegraphics[scale=0.4]{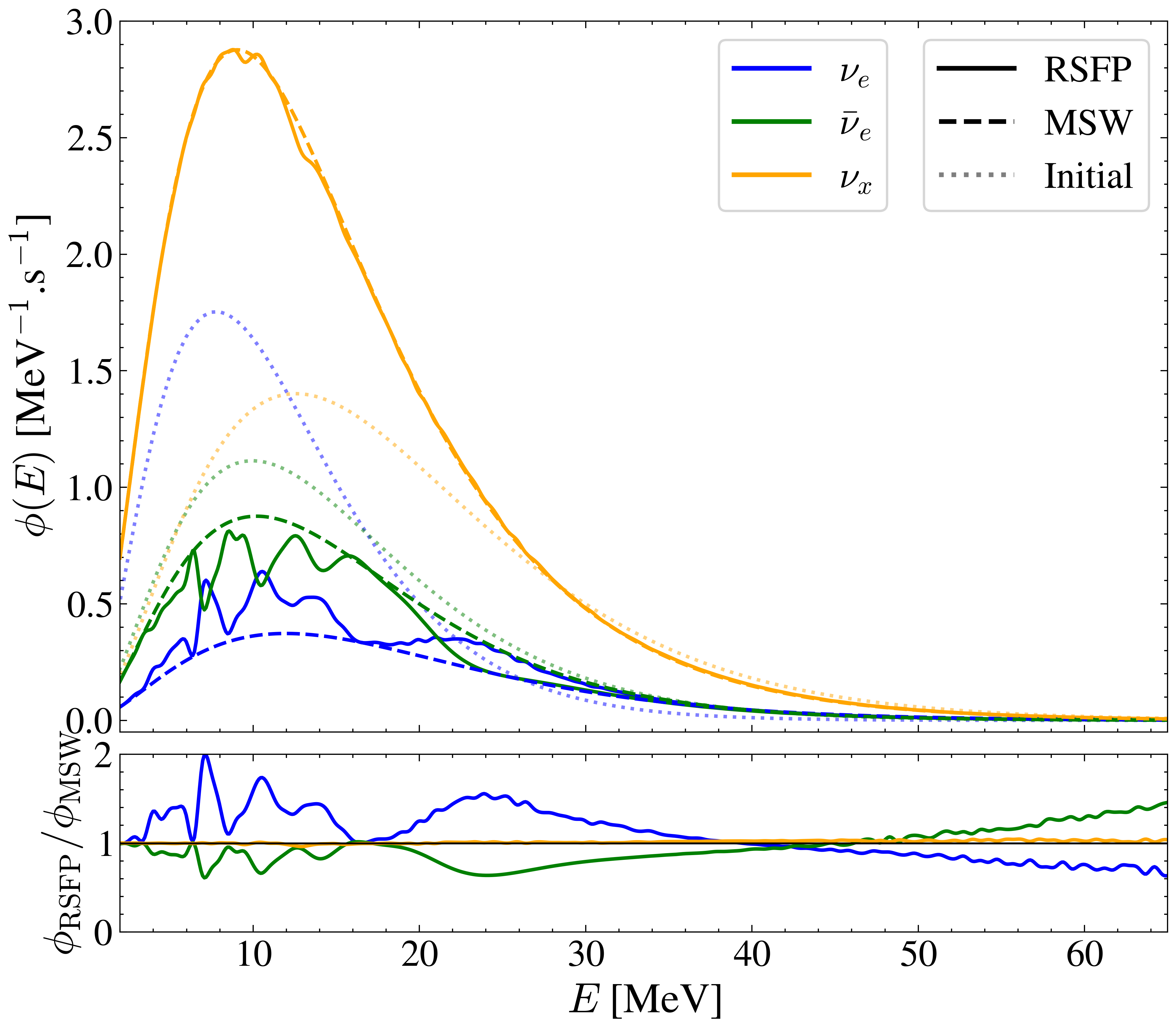} 
\includegraphics[scale=0.4]{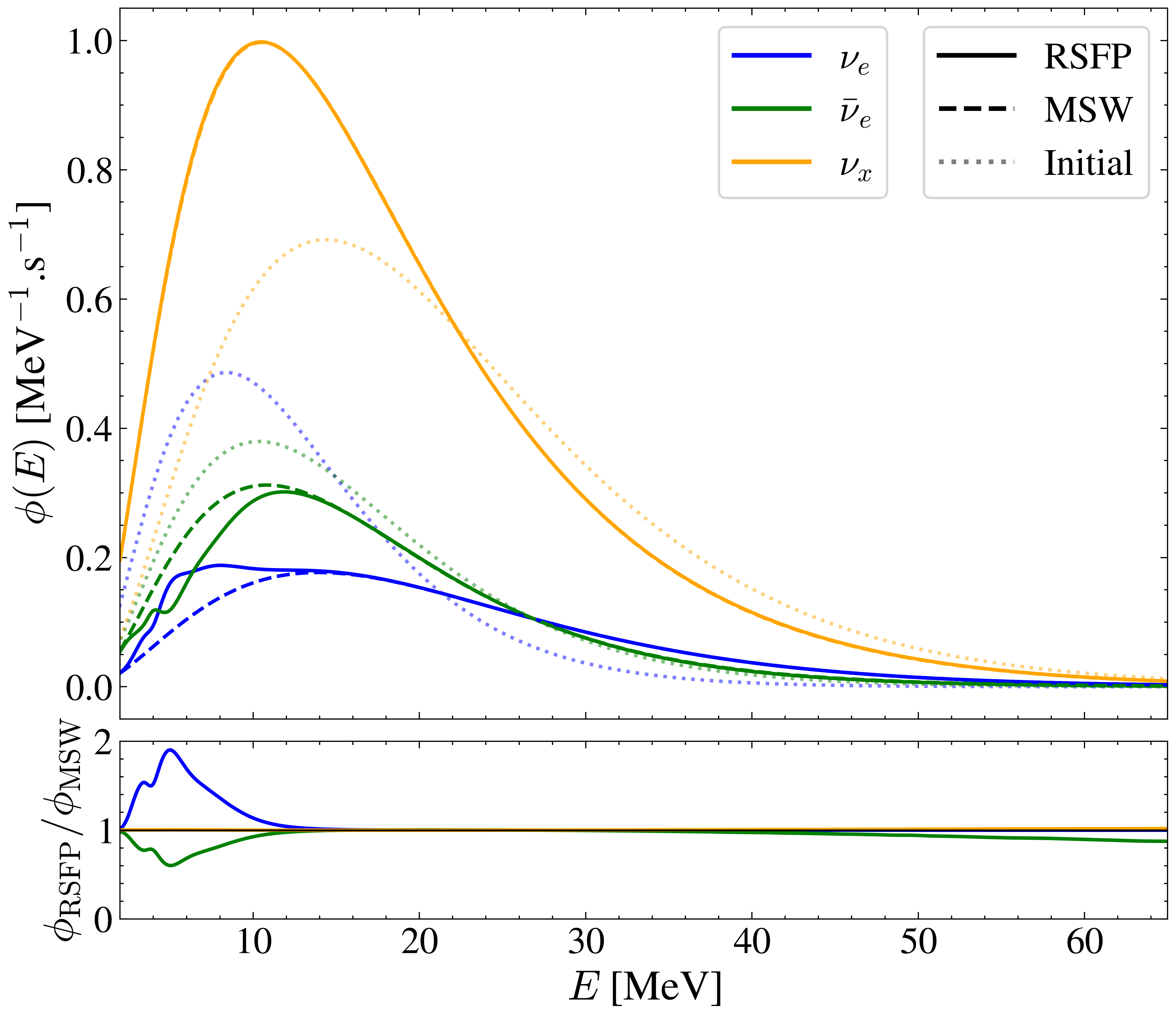} 
 \end{center}
 \caption{Neutrino fluxes 
 as a function of neutrino energy, in presence of a non-zero magnetic moment. The results are for Majorana neutrinos and NO. The panels correspond to the three different time snapshots of 0.02 s (upper left), 0.13 s (upper right) and 2 s (bottom). The results, obtained in a full 3$\nu$ framework, are for the neutrino-driven supernova A17 described in section~\ref{sec:A17}. For the each time the corresponding magnetic field profile is shown in figure~\ref{fig:A17mag}. 
 In each panel the upper figure shows the $\nu_e$, $\bar{\nu}_e$ and $\nu_x$ fluxes. Note that the non-electron neutrino flux noted as $\nu_x$ here is the summed flux over all non-electron flavors. The cases shown correspond to i) no flavor evolution ("Initial"); ii) MSW effect only ("MSW"); iii) spin-flavor resonances and the MSW effect ("RSFP"). The value of the transition magnetic moments here is $\mu = 10^{-14}~\mu_B $.
 Below each of the flux figures, the ratio of the fluxes with and without the inclusion of the neutrino magnetic moment is also given.}\label{fig:A17flux}
\end{figure}

Using a 3$\nu$ flavor framework, we have determined the neutrino evolution for different values of the neutrino magnetic moment. The computations are based on the detailed inputs shown in figures \ref{fig:A17}-\ref{fig:A17mag}. 
Figure \ref{fig:A17flux} shows the results for the neutrino fluxes including a neutrino magnetic moment of $\mu = 10^{-14}~ \mu_B $. Sizable effects are found in NO; whereas for IO the neutrino magnetic moment has a very small impact, independently of the considered time snapshot. This is due to the non-adiabaticity of neutrino evolution through the spin-flavor resonances. Similarly, for $\mu = 10^{-15} \mu_B $ and NO, we
do not find sizable effects.

\begin{figure}
\begin{center}
\includegraphics[scale=0.4]{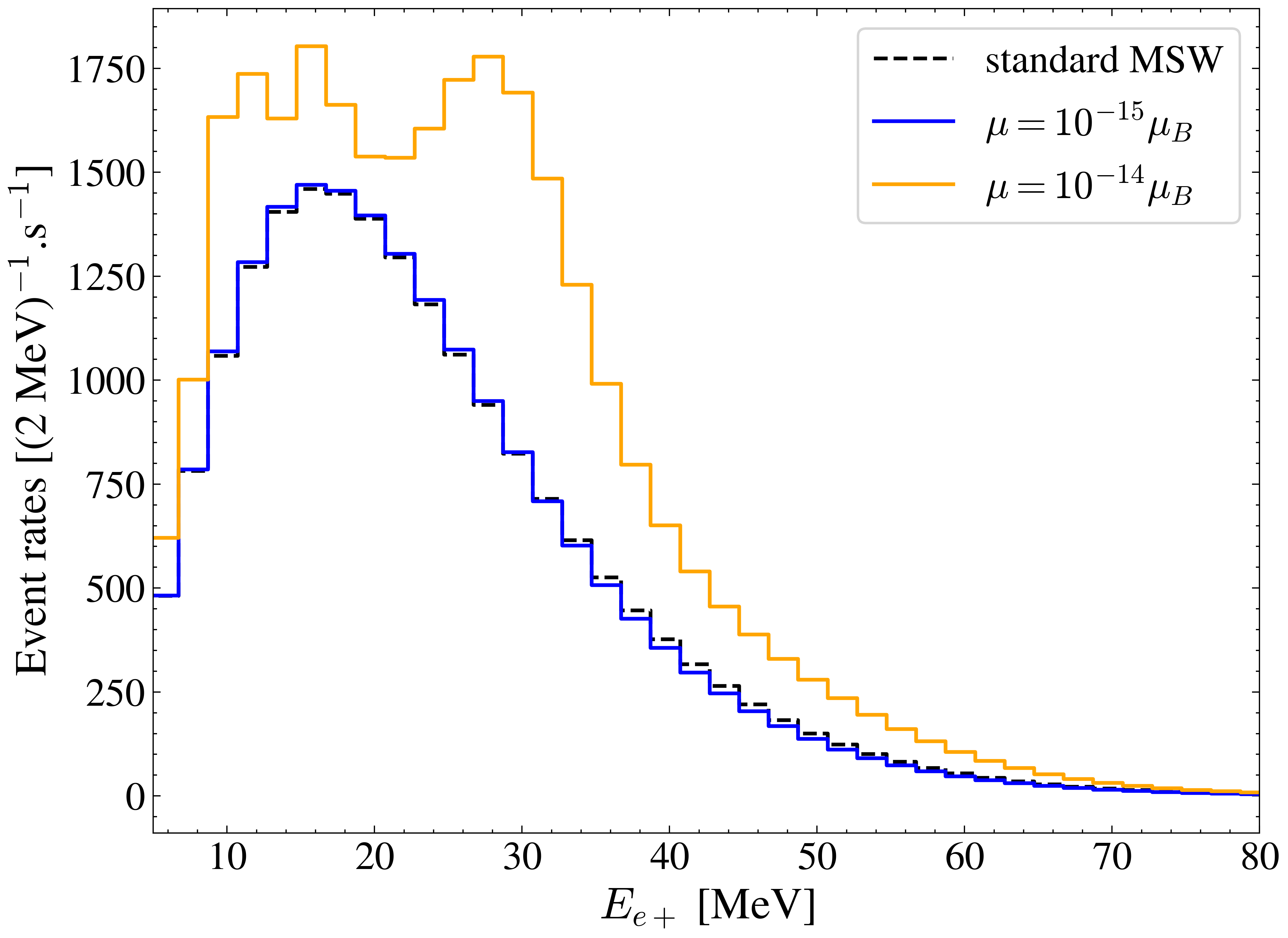} 
\includegraphics[scale=0.4]{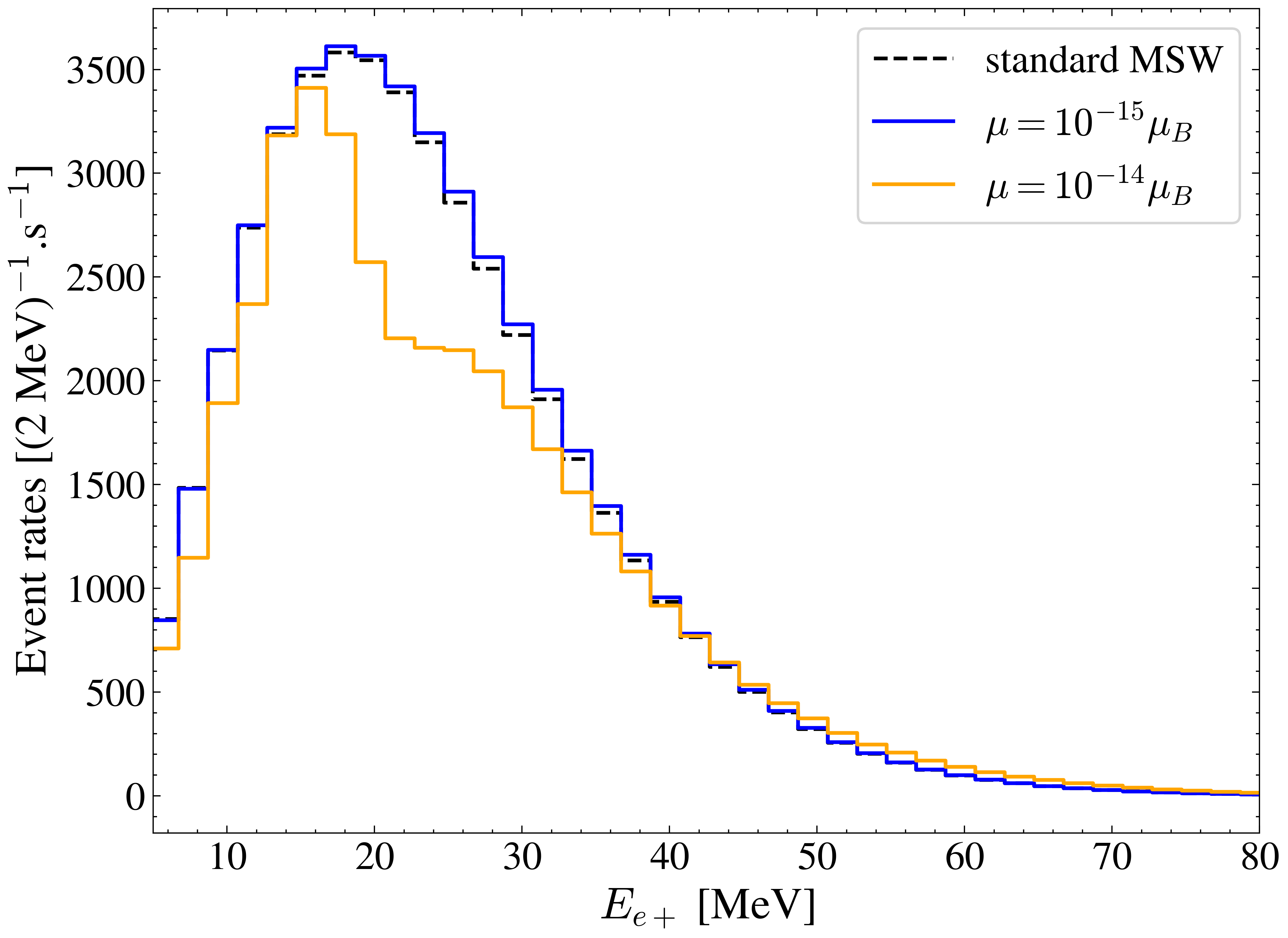} 
\includegraphics[scale=0.4]{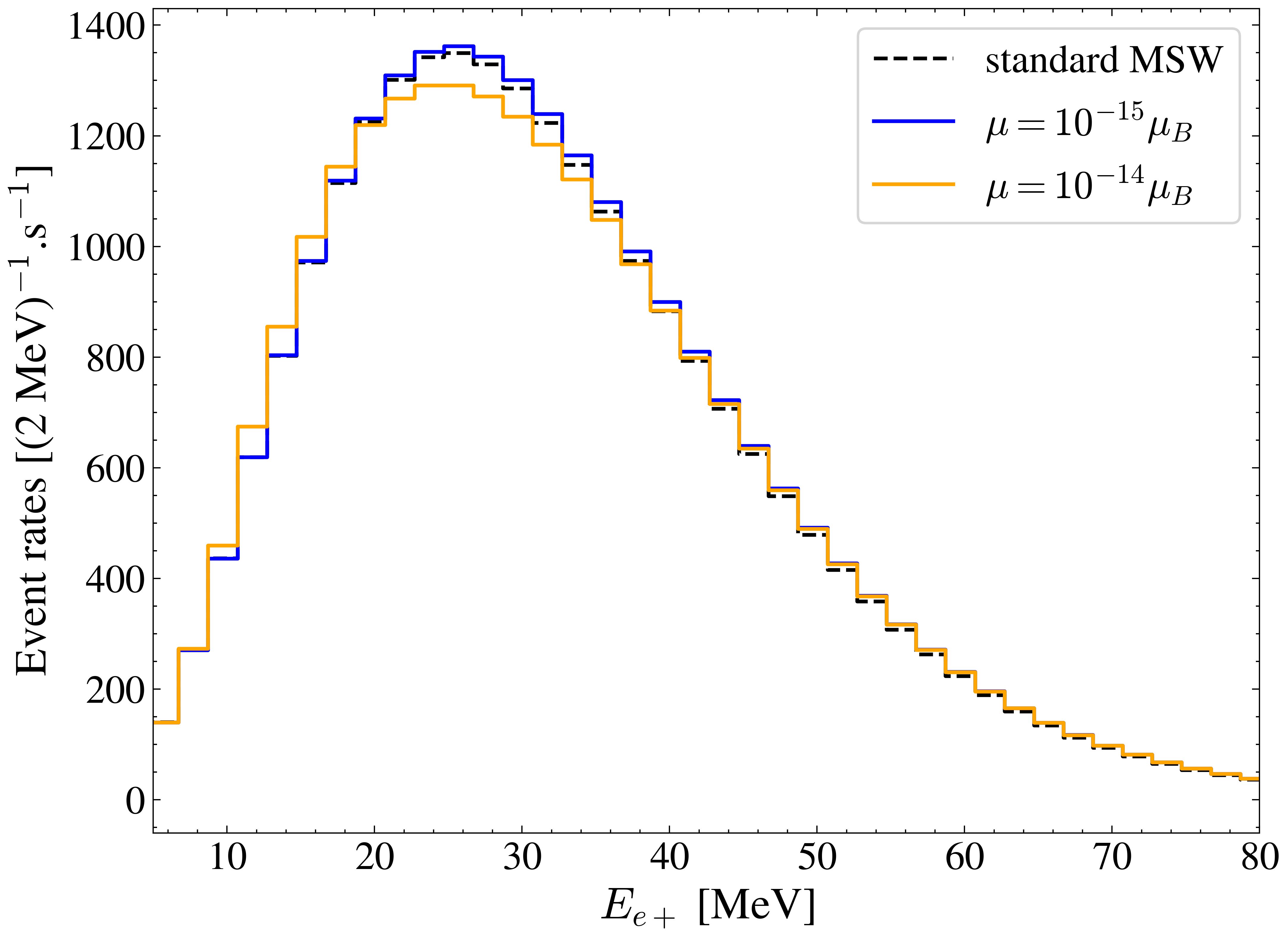} 
 \end{center}
 \caption{Event rate associated to inverse-beta decay in the HK detector, as a function of positron energy, including a non-zero neutrino magnetic moment for Majorana neutrinos. The results shown, based on the three-dimensional neutrino-driven supernova model A17, correspond to the three representative times of 0.02 s for the neutronization burst and NO (upper left), 0.13 s for the accretion phase and NO (upper right) and 2 s for the cooling of the newly formed proton-neutron star and IO (bottom). In each figure,  the predictions are in absence of the neutrino magnetic moment ("standard MSW"), or including its coupling to the supernova magnetic field for the two values of the transition magnetic moments of $10^{-15} \mu_B$ and $10^{-14} \mu_B$.}
 \label{fig:A17ev}
\end{figure}

Figure \ref{fig:A17ev} presents the results on the event rates as a function of positron energy, for the HK detector and for the three time snapshots corresponding to the three phases of the neutrino emission. The impact of the neutrino magnetic moment on the event rate reflects the one on the neutrino fluxes (figure \ref{fig:A17flux}), also taking into account the cross sections dependence on $E^2$. 
The results show that the neutrino magnetic moment produces a significant impact on the event rate for the neutronization and the accretion phase, for NO and values of the transition magnetic moments $\mu = 10^{-14}~\mu_B $, well below the current bounds (table \ref{tab:mulimits}). On the contrary we find that the neutrino magnetic moment has a small impact at the late times. 
For the IO case, we find effects that are small even at early times (figures not shown).
One can also see that for $\mu \lesssim 10^{-15}~\mu_B $ the results are very close to the ones including the MSW effect only for both NO and IO. 

\subsection{A26 model: neutrino fluxes and event rates}
Here we discuss the impact of the neutrino magnetic moment when inputs on the neutrino fluxes, angle-averaged matter and magnetic radial field profiles are taken for the magnetorotational supernova model A26.

\begin{figure}
\begin{center}
\includegraphics[scale=0.45]{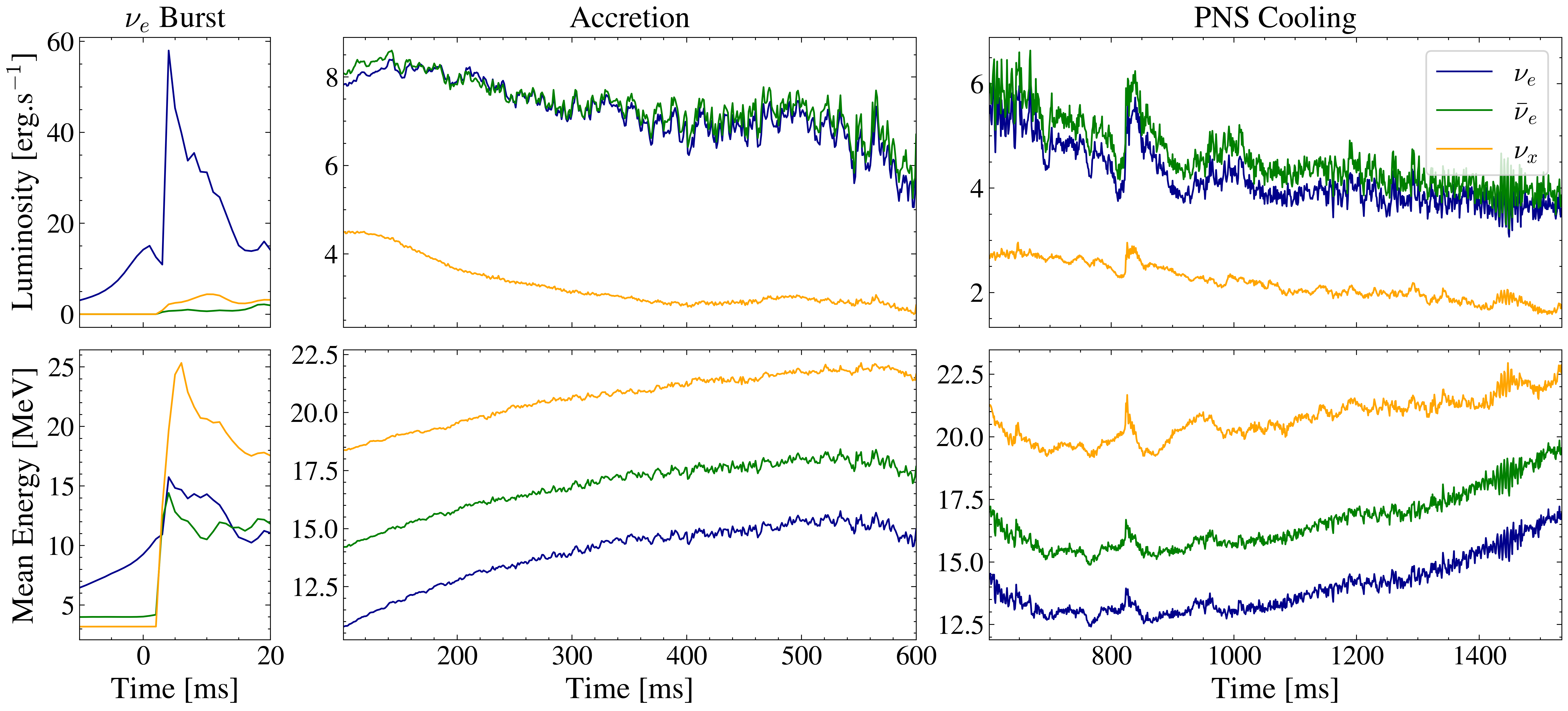} 
 \end{center}
 \caption{Model A26: Neutrino luminosities ($\times 10^{52}$) 
 and mean energies (lower figures) as a function of time for a magnetorotational supernova with a $26\,{\rm M}_{\odot}$ progenitor forming a black-hole (section \ref{sec:A26}). 
  The three figures correspond to the neutronization burst (left),
  accretion phase (middle) and later times just before black hole formation (right) at $t = 1.5\,$s (corresponding to 1.3\,s post-bounce). The results shown are at $r = 500 ~ {\rm km}$ from the supernova center.}\label{fig:A26}
\end{figure}

 The neutrino luminosities and mean energies from the A26 simulations are given in figure~\ref{fig:A26} considering three time windows. By comparing the neutrino parameters, one can see that the luminosities and the mean energies are higher in the A26 model than in the neutrino-driven model A17 (figure~\ref{fig:A17}). As for the A17 model, the mean energies follow
a clear hierarchy with $\langle E_{\nu_e} \rangle < \langle E_{\bar{\nu}_e} \rangle < \langle E_{\nu_x} \rangle$, both during the accretion phase and at late times.
One can also see a steady increase of the neutrino mean energies due to the compression of
baryonic matter, before black hole formation.
 
\begin{figure}
\begin{center}
\includegraphics[scale=0.5]{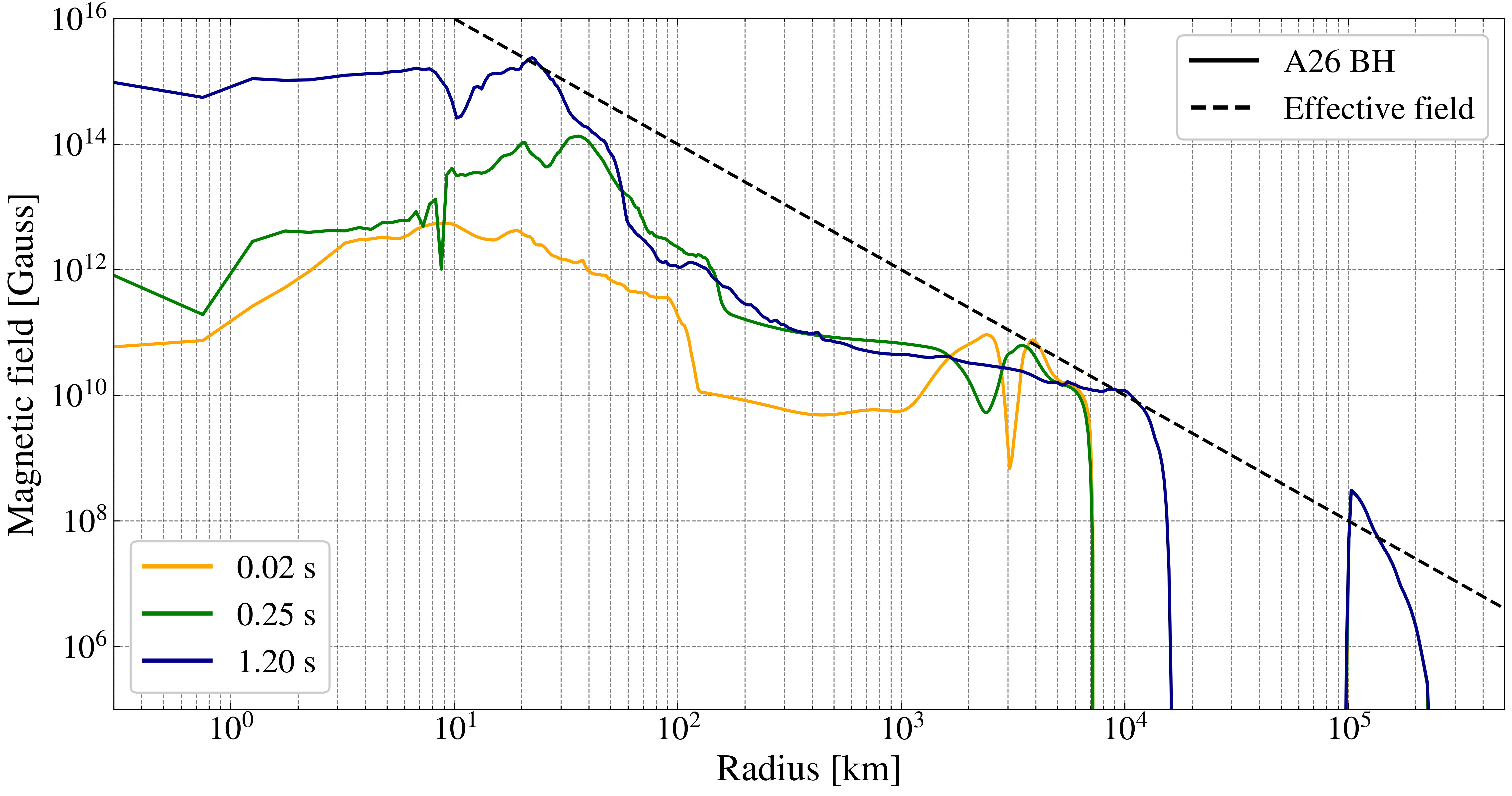} 
 \end{center}
 \caption{Model A26: Angle-averaged transverse magnetic field strength profiles as a function of the distance from the three-dimensional magnetorotational supernova model. Three different times are shown, namely 0.02 ${\rm s}$ for the neutronization burst, 0.25 ${\rm s}$ for the accretion phase and 1.20 ${\rm s}$ before black-hole formation. For comparison, the dashed line indicates the power law magnetic field profile used in the schematic model (see section \ref{Sect:schematic}).}\label{fig:A26mag}
\end{figure}

 The magnetic field profiles used in our calculations are presented in figure \ref{fig:A26mag}, along with the power-law radial profile \eqref{eq:Bprof}, used in the schematic model (section \ref{Sect:schematic}). Three different time snapshots are taken for the magnetic field, as representative of the radial dependence of the transverse magnetic field strength, for the three time windows shown in figure \ref{fig:A26}. 

\begin{figure}
\begin{center}
\includegraphics[scale=0.5]{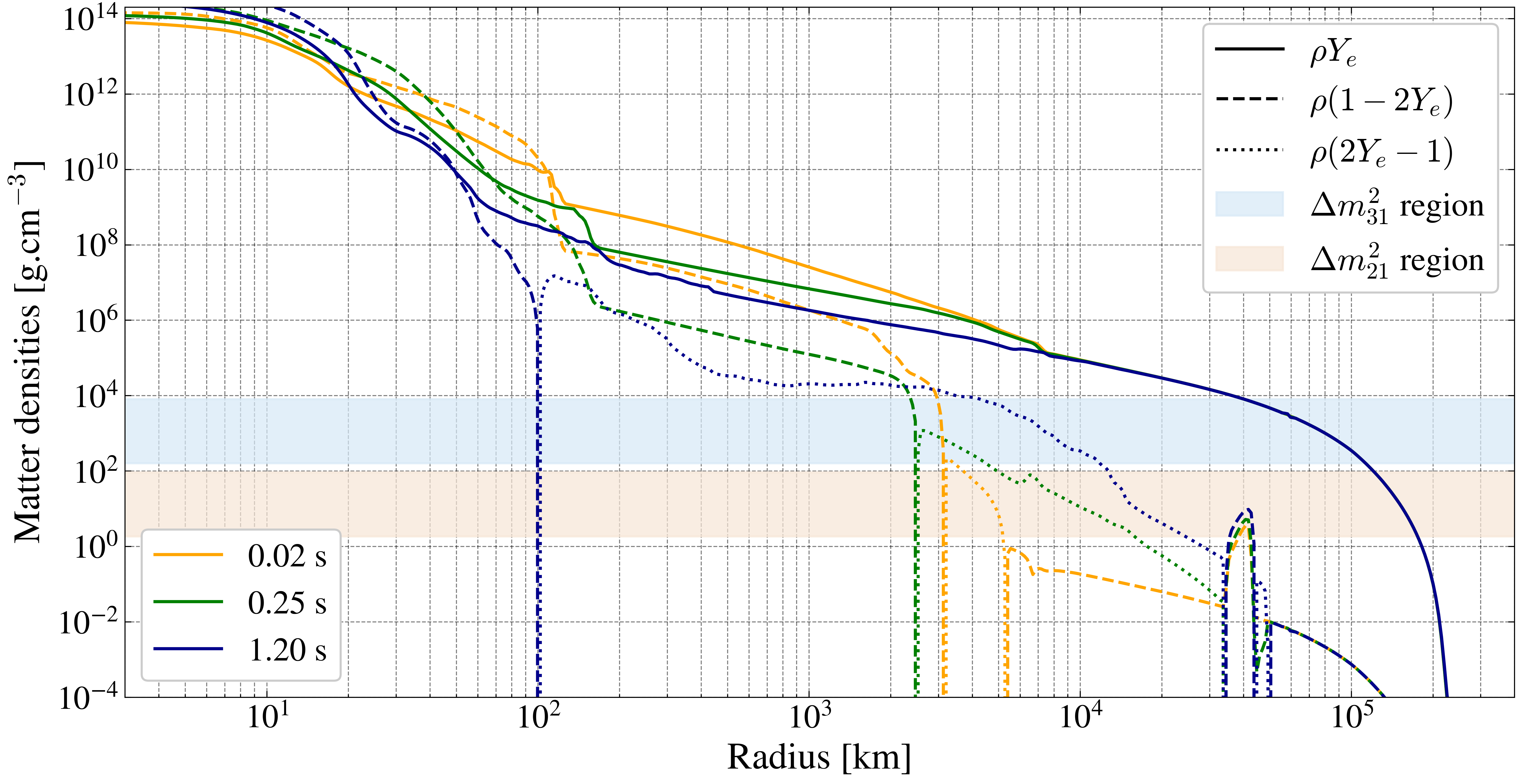} 
 \end{center}
 \caption{Model A26: Matter density as a function of radius for three time snapshots, at early times, during the accretion phase and late times, as in figure \ref{fig:A26mag}. The results are for the three-dimensional magnetorotational model A26 leaving a black-hole \cite{ObergaulingerAloy:2021PaperIII}. The {\it l.h.s.} of the RSFP-H \eqref{eq:RSFP-H} and RSFP-L \eqref{eq:RSFP-L} resonance conditions are also shown. The colored bands, corresponding to a range of neutrino energies, show the approximate locations where both the MSW and the RSFP-H and RSFP-L resonances can be fulfilled.} \label{fig:A26res}
\end{figure}

As in the case of model A17 (figure~\ref{fig:A17res}), we present, in figure~\ref{fig:A26res}, the matter density profile $\rho Y_e$, as a function of radius, for three time snapshots in figure~\ref{fig:A26mag}. 
In order to show the (approximate) location of the spin-flavor resonance, the {\it l.h.s.} of the RSFP-H \eqref{eq:RSFP-H} and RSFP-L \eqref{eq:RSFP-L} resonance conditions are also presented with bands corresponding to the {\it r.h.s.} of the same relations for $E \in [2, 100]$ MeV. The $\pm \rho (2 Y_e - 1) $ are also displayed. 

It is to be noted that the RSFP-L condition appears also to be fulfilled in the magnetic field gap for $10^{4}~{\rm km} < r < 10^{5} ~{\rm km}$. As discussed in section~\ref{sec:A17-results}, the apparently unmagnetized layers in the progenitor result from the one-dimensional prescription used to model magnetic-field transport and flux expulsion during the advanced stages of stellar evolution. In model A26, the outer magnetic-field gap coincides with an oxygen-rich shell. Three-dimensional MHD simulations have shown that convection can drive efficient small-scale dynamo amplification in oxygen-burning shells, producing mean magnetic fields of order $10^{10}$--$10^{11}\,{\rm G}$ \cite{Varma:2021zqg,Varma:2023kum,Leidi:2023MHD}. Moreover, Griffiths et al.~\cite{Griffiths_II_2026arXiv260522938} find that convective amplification can magnetically connect regions that remain disconnected in one-dimensional progenitor models. The field-free gap in A26 should therefore be regarded as a consequence of the one-dimensional prescription rather than as a robust physical prediction. Existing simulations, however, do not yet demonstrate that this particular region would be completely or uniformly magnetized.

\begin{figure}
\begin{center}
\includegraphics[scale=0.4]{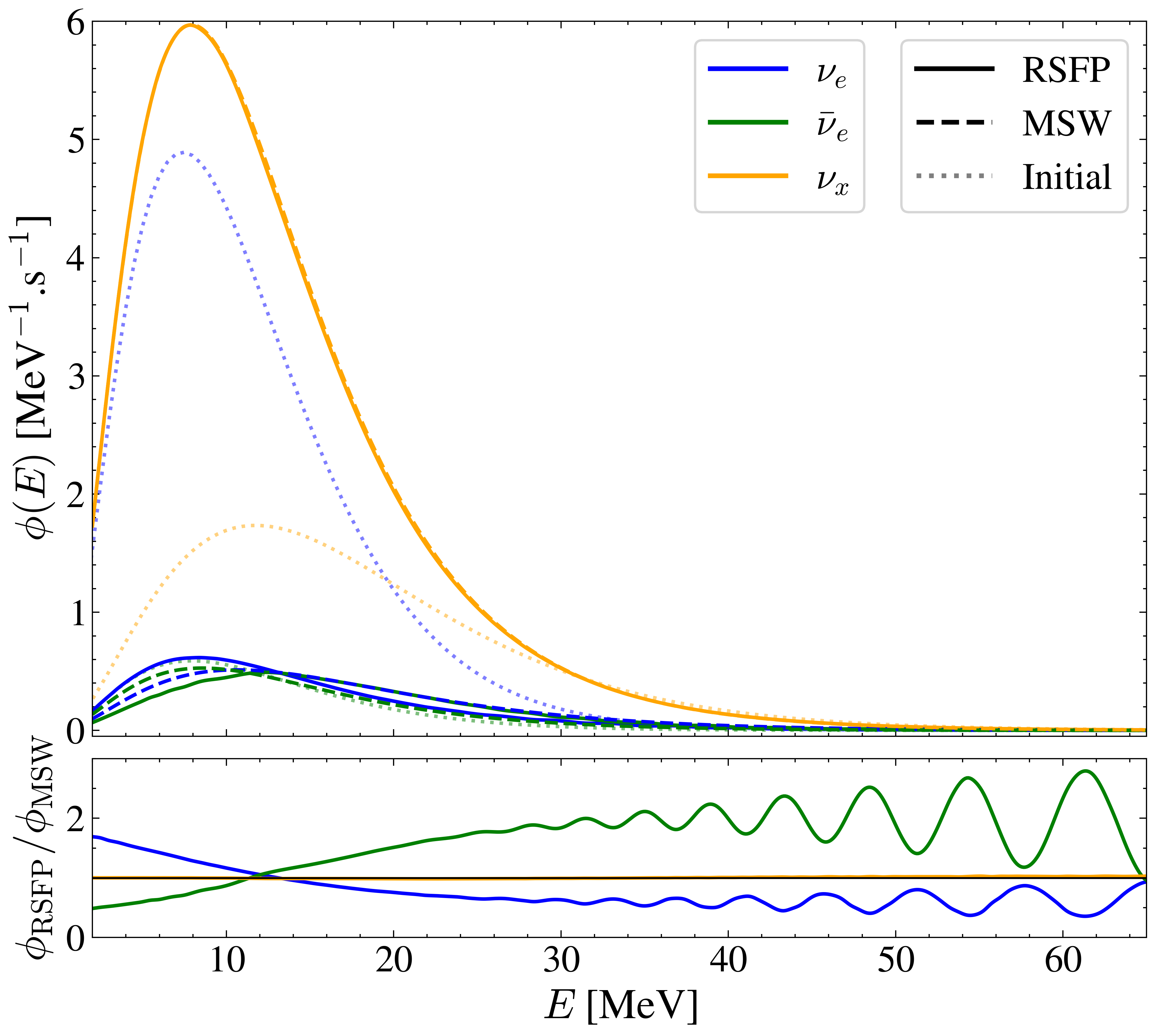} 
\includegraphics[scale=0.4]{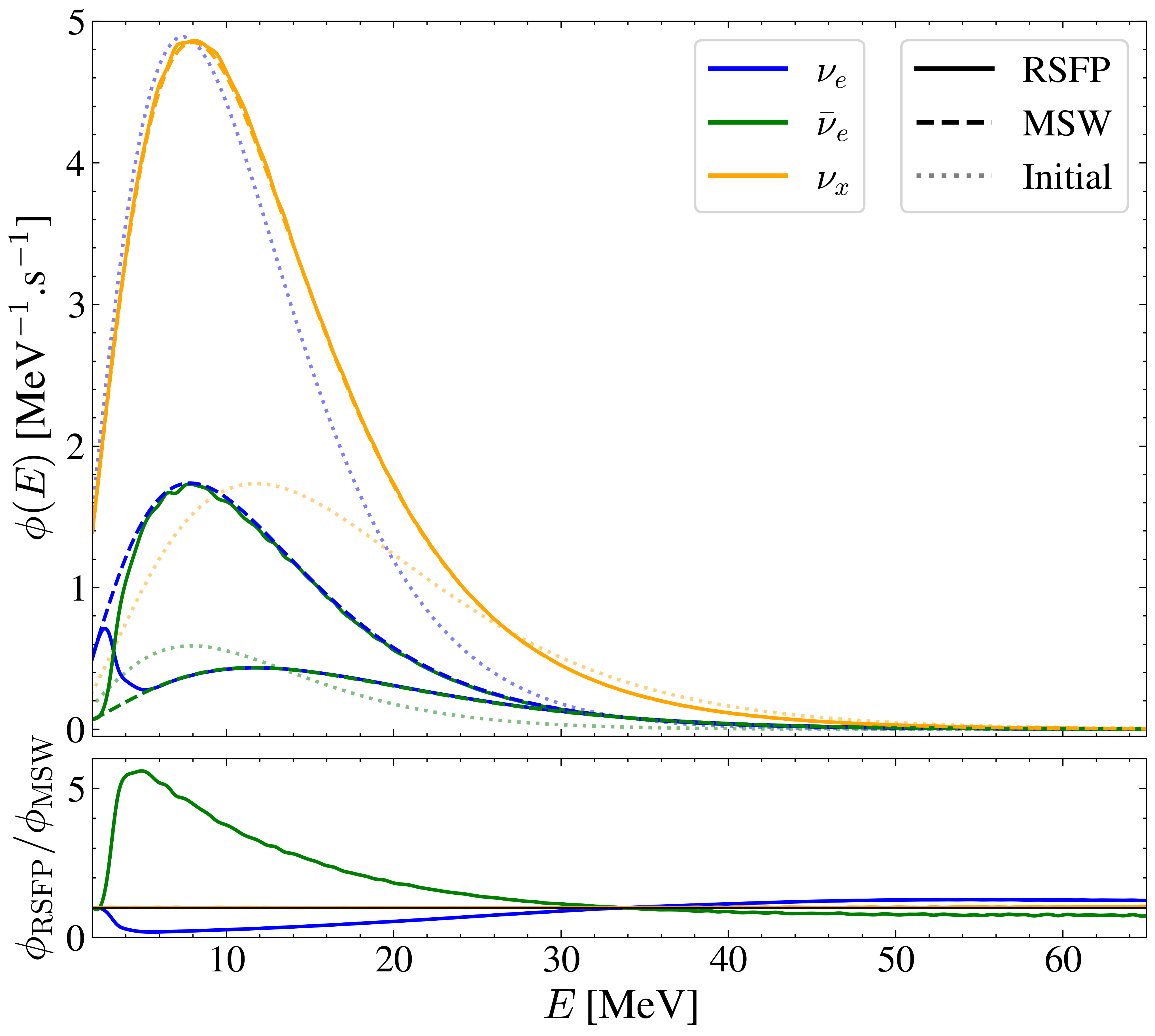} 
\includegraphics[scale=0.4]{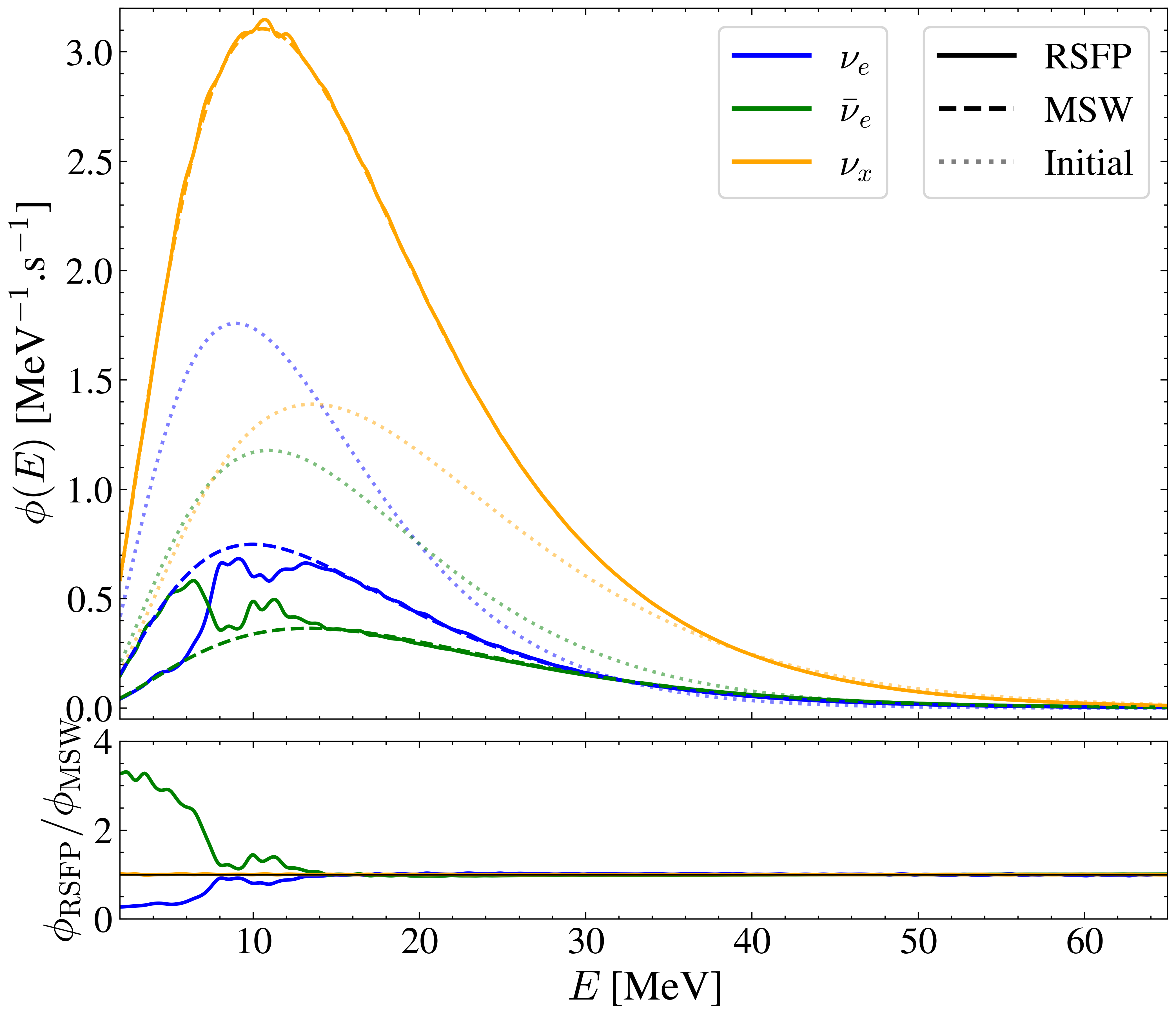} 
 \end{center}
 \caption{Neutrino fluxes, as a function of neutrino energy. The model is the magnetorotational supernova A26 described in section~\ref{sec:A26}. 
 In each figure the $\nu_e$, $\bar{\nu}_e$ and $\nu_x$ fluxes are given. Note that the non-electron
neutrino flux noted as $\nu_x$ here is the summed flux over all non-electron flavors. In each panel the results, obtained in a full 3$\nu$ framework, correspond to: i) no flavor evolution ("Initial"); ii) only considering the MSW effect ("MSW"); iii) including both the spin-flavor resonances and the MSW effect ("RSFP"). Below each of the figures the ratio of the fluxes with spin-flavor resonances and without them is given. The panels correspond to the following cases. Upper left: NO, $10^{-14} \mu_B$, $t = 0.02$ s.
 Upper right: IO, $10^{-14} \mu_B$, $t = 0.02$ s. Lower panel: IO, $10^{-14} \mu_B$, $t = 0.25$ s.}\label{fig:A26flux}
\end{figure}

The impact of the neutrino magnetic moment on the $\nu_e, \bar{\nu}_e$ and $\nu_x$ fluxes is shown in figure \ref{fig:A26flux}. The results shown correspond to the early times and a transition magnetic moment of $10^{-14} \mu_B$, for either the NO or the IO cases. Note that for IO, we find a modification of the $\bar{\nu}_e$ fluxes below 20 MeV by up to a factor of 2 also for $\mu = 10^{-15} \mu_B$. 
Indeed, 
in the A26 model, the magnetic fields in the resonance regions are larger by one order of magnitude (or more), 
as can be seen by comparing 
figures~\ref{fig:A17mag}-\ref{fig:A17res} with figures~\ref{fig:A26mag}-\ref{fig:A26res}. This influences the adiabaticity of the evolution at the resonances.

\begin{figure}
\begin{center}
\includegraphics[scale=0.4]{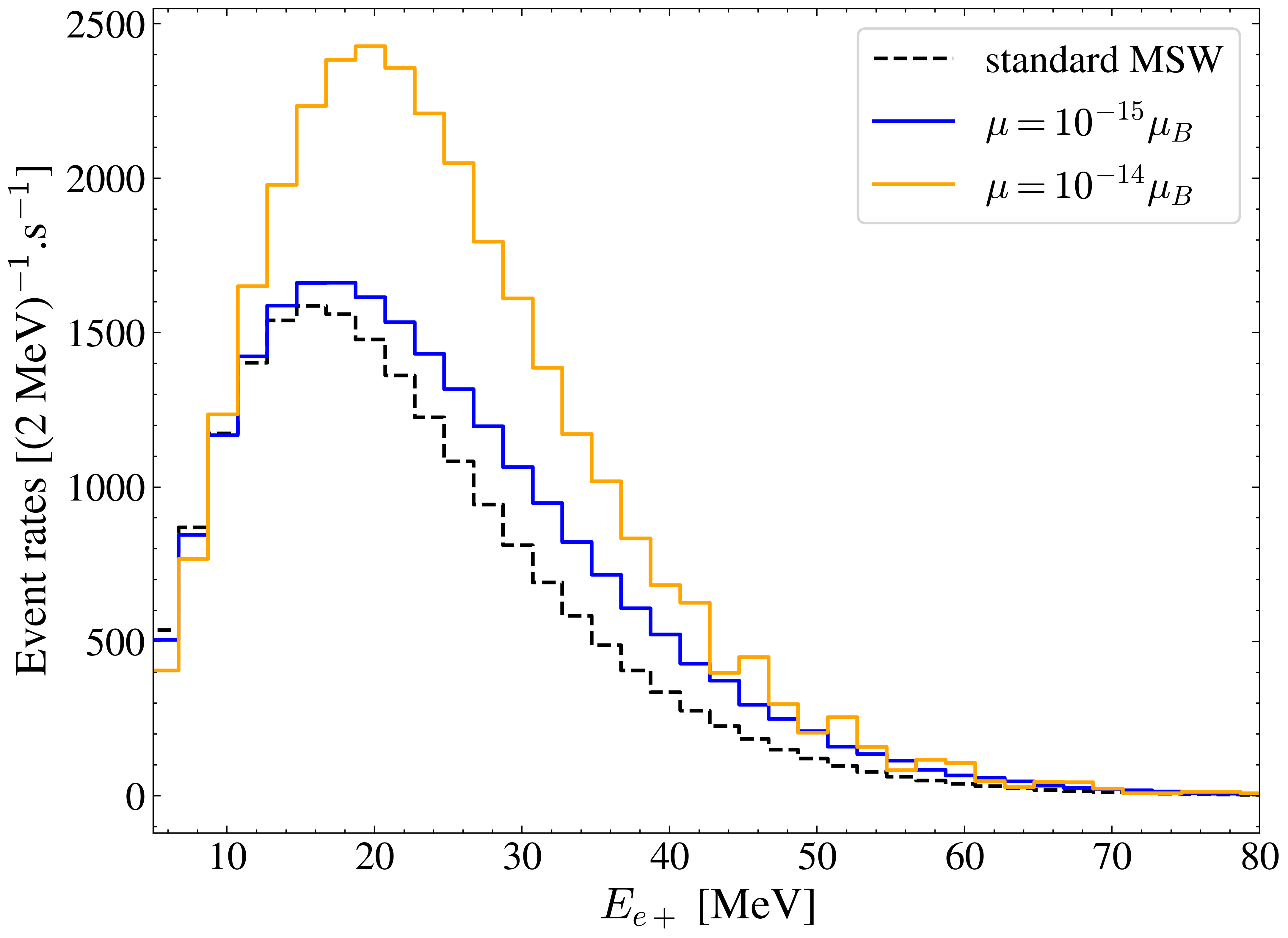} 
\includegraphics[scale=0.4]{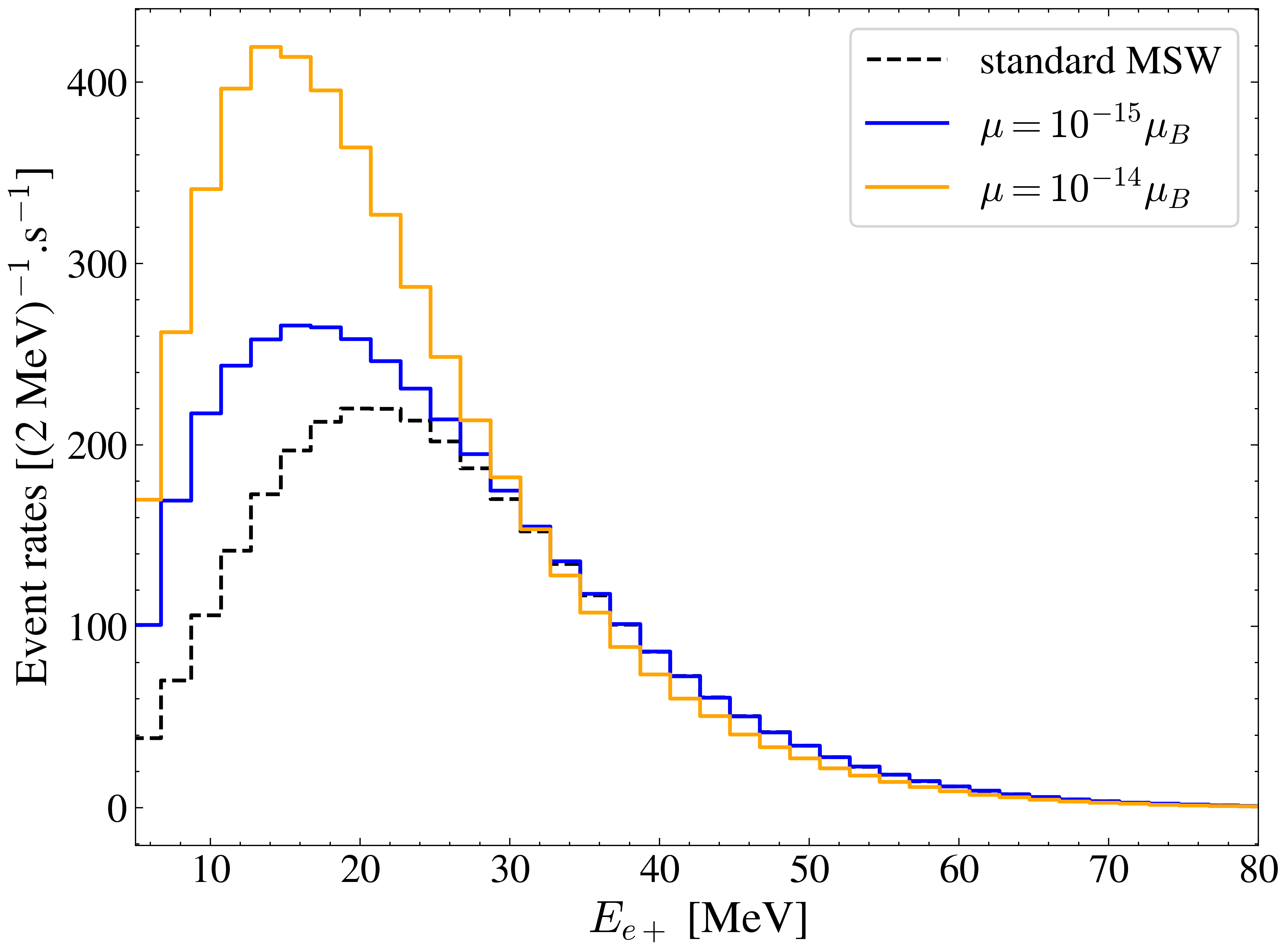} 
\includegraphics[scale=0.4]{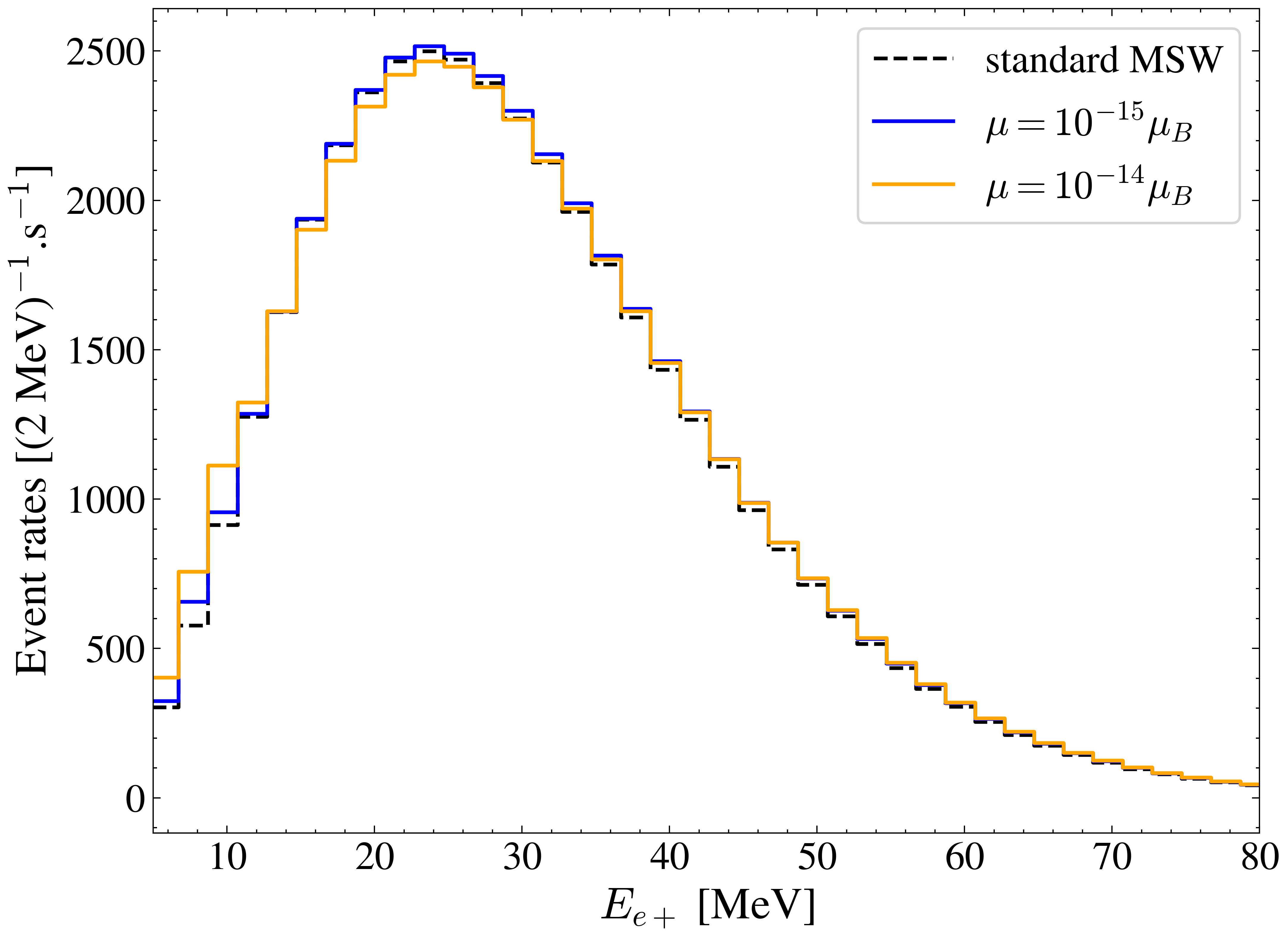} 
 \end{center}
 \caption{Event rate associated with inverse-beta decay as a function of positron energy for the magnetorotational supernova model A26, with a black-hole remnant. The galactic supernova location is 10 kpc from the Earth. The results include a non-zero neutrino magnetic moment and are for Majorana neutrinos. The figures correspond to 0.02 s, HK and NO (upper left); 0.02 s, JUNO and IO (upper right) and 0.25 s, HK and IO (lower figure).  
 The predictions are in absence of the neutrino magnetic moment ("standard MSW") or including its coupling to the supernova magnetic field for the two values of the transition magnetic moments of $10^{-15} \mu_B$ and $10^{-14} \mu_B$.}
 \label{fig:A26ev}
\end{figure}

The predicted event rate in the JUNO\footnote{The results for NO can be obtained by scaling.} and HK detectors are shown in figure~\ref{fig:A26ev}.
For the late times, no significant effect is found for both neutrino mass orderings. On the contrary, we find that the impact of RSFP on the neutrino fluxes is large, even for transition magnetic moments as small as $10^{-15} \mu_B$ during the neutronization burst and the accretion phase.

\section{Conclusions}
\label{Sect:conclusions}

In this work we have investigated the impact of the neutrino magnetic moment for Majorana neutrinos on the neutrino signal from a future galactic supernova. The study comprised two complementary parts. In the first part we determined the effects of a non-zero magnetic moment via a schematic model.
In the second part, which is the main focus of our work, we anchored our computations on detailed multidimensional supernova simulations. 

The schematic model, often employed in the literature to explore the effects from the neutrino magnetic moment effects takes a power-law dependence for the radial matter and transverse magnetic field profiles. 
We presented both the neutrino matter eigenvalues and the survival probabilities identifying the spin-flavor resonances, due to the neutrino coupling to the magnetic field, obtained by solving numerically the neutrino evolution for 3$\nu$ flavors. Thus, our results go beyond the analytical approximations used so far to determine the neutrino evolution at the resonances in the schematic model. The results for the fluxes and event rates present significant effects for NO with $\mu {\rm B}_0 > 200 ~\mu_B {\rm G}$, and  for IO with $\mu {\rm B}_0 = 100 ~\mu_B {\rm G}$.    

In the second part, we numerically determined the neutrino evolution by solving the equations with a non-zero transition magnetic moment, exploiting information from detailed three-dimensional simulations for two models. These were a neutrino-driven supernova with zero-age main sequence mass of 17 $M_{\odot}$ leaving a neutron-star, and a magnetorotational driven supernova with zero-age main sequence mass of 26 $M_{\odot}$ leaving a black-hole remnant. Detailed inputs from the simulations were used, both for the neutrino fluxes, the matter and for the magnetic field profiles. In our time-dependent analysis we used three snapshots as representative of the different supernova neutrino emission phases. 

For A17, we found that, if a galactic supernova explodes at 10 kpc from the Earth, neutrino transition magnetic moments with $\mu \ge 10^{-14} \mu_B $ and NO could produce significant modifications of the $\bar{\nu}_e$ fluxes, as well as of the IBD event rate in the running JUNO and upcoming HK detectors. This is true for both the neutronization and the accretion phase, whereas for the cooling phase effects are tiny.
For IO we do not find a sizable impact of $\mu$. 
For the model A26, our results for the early phase ($\simeq$ 20 ms) show that the effects from the spin-flavor resonances due to $\mu \ge 10^{-14} \mu_B $ in NO and $\mu \ge 10^{-15} \mu_B $ in IO can be important. 

Further investigations beyond the current work can be envisaged, that would e.g. using angle-dependent magnetic field profiles, thus introducing angle-dependent magnetic field effects, as done in ref.~\cite{Manno:2026ikk}. The implementation of effects coming from the $\nu\nu$ interaction is also beyond the scope of the present study. 

Although magnetorotational supernovae are expected to be rare, likely a few percent of the overall supernova population, the lucky occurrence of
such an event could provide crucial information. Important clues on the presence of strong magnetic fields and that the explosion was magnetorotational could come 
by independent multimessenger observations. Relevant diagnostics include a bolometric light curve and photospheric-velocity evolution consistent with sustained spin-down power from a newly born millisecond magnetar \cite{Kasen:2010Magnetar}; an associated long-duration gamma-ray burst, or radio and broadband emission whose spectrum and evolution require a collimated mildly or ultra-relativistic outflow \cite{Soderberg:2010Engine}; pronounced axisymmetric or bipolar ejecta inferred from spectropolarimetry or nebular emission-line profiles \cite{Maeda:2008Asphericity}; and a gravitational-wave signal whose waveform is statistically favoured by magnetorotational rather than neutrino-driven explosion templates \cite{Powell:2024GW}. Complementary evidence could be provided by a strong $r$-process extending beyond the second abundance peak, as predicted for favourable magnetic-field configurations in some magnetorotational models \cite{Winteler:2012hu,Reichert:2024vyd}. Possible observational routes include late-time near- and mid-infrared nebular lines attributable to freshly synthesized heavy elements and isotope-specific radioactive-decay gamma-ray lines from the supernova or its remnant \cite{Ricigliano:2025Rprocess,Liu:2025GammaRays}. For an A26-like event, an abrupt termination of the MeV neutrino signal would additionally support the formation of a black-hole \cite{Beacom:2000qy,Sumiyoshi:2007BH}, although this would diagnose the compact remnant rather than uniquely identify the explosion mechanism. Conversely, sustained magnetar-powered emission would favour a long-lived neutron-star remnant and would therefore support the broader magnetorotational scenario while disfavouring the specific black-hole-forming outcome of A26. Although none of these observables would be unambiguous on its own, the joint detection of several mutually consistent and independent diagnostics would constitute compelling evidence in favour of a magnetorotational explosion.

The results of the present work, for Majorana neutrinos, show that a supernova event of this class, or a neutrino-driven supernova with strong magnetic fields could provide unique sensitivity to neutrino transition magnetic dipole moments.  If IO turns out to be the mass ordering of neutrinos the neutrino fluxes appear modified for $\mu \ge 10^{-15}~\mu_B$, well below current bounds. In case NO is the true mass ordering, spectral modifications like the ones discussed in the present study could be present for $\mu \ge 10^{-14} \mu_B$. Thus, based on the model-independent bounds for Dirac neutrinos of ref.~\cite{Bell:2005kz}, one could maybe get indications that neutrinos are Majorana particles by combining neutrino and multimessenger observations from a future supernova.

Our investigation of the neutrino magnetic moment impact using detailed supernova neutrino-driven and  magnetorotational supernova simulations, both including
strong
magnetic fields, is  encouraging. Significant effects are found even for neutrino magnetic moments well below current bounds. Clearly, the next galactic or extragalactic core-collapse supernova
will bring crucial information for confirming our current understanding of the supernova mechanism(s) and unique information on still unknown neutrino properties like the neutrino magnetic moment, and maybe, the neutrino nature.

\acknowledgments 
M. Cristina Volpe would like to thank
Matteo Bugli for useful discussions and information. C. Ehrhardt thanks the Astroparticle and Cosmology (APC) Laboratory for their hospitality and support during the project. This work was supported by "CNRS Nucléaire et Particules" through the "NUFRONT" Theory Masterproject and ANR-18-IDEX-0001.
MAA and MO acknowledge support from grants PID2021-127495NB-I00 and PID2025-171322NB-C22, funded by MCIN/AEI/10.13039/501100011033 and by the European Union “NextGenerationEU". They also acknowledge support from the Prometeo excellence programme grant CIPROM/2022/13 funded by the Generalitat Valenciana. The supernova computations have been performed on the Red Española de Supercomputación (RES) on MareNostrum (grants AECT-2025-1-0012 and AECT-2025-2-0006).

\bibliographystyle{JHEP}
\bibliography{references_modif}

\end{document}